\documentclass[10pt, a4paper, conference]{IEEEtran}

\usepackage[linesnumbered, ruled, vlined]{algorithm2e}

\usepackage{amsbsy,amsmath,amssymb,epsfig,bbm,mathrsfs,multirow,amsthm}
\usepackage{float}
\usepackage{cite}
\usepackage{array, multirow, graphicx}
\usepackage{graphics}
\usepackage{comment}
\usepackage{algpseudocode}
\usepackage{hyperref} 
\usepackage{color}
\usepackage{lipsum}
\usepackage{caption}
\usepackage{subcaption}
\usepackage{flushend}

\usepackage{multirow} 
\usepackage{url} 
\usepackage{hyperref}

\usepackage{setspace} 

\begin{document}


\title{Economic Systems in Metaverse: Basics, State of the Art, and Challenges}

\author{
 \IEEEauthorblockN{
      Huawei~Huang\IEEEauthorrefmark{1},
      Qinnan~Zhang\IEEEauthorrefmark{2},
      Taotao~Li\IEEEauthorrefmark{1},
      Qinglin~Yang\IEEEauthorrefmark{1},
      Zhaokang~Yin\IEEEauthorrefmark{1},
      Junhao~Wu\IEEEauthorrefmark{1},\\
      Zehui~Xiong\IEEEauthorrefmark{3},
      Jianming~Zhu\IEEEauthorrefmark{2},
      Jiajing~Wu\IEEEauthorrefmark{1},
      Zibin~Zheng\IEEEauthorrefmark{1}
      }
    \IEEEauthorblockN{
    \IEEEauthorrefmark{1}{Sun Yat-Sen University, China. Email: huanghw28@mail.sysu.edu.cn}\\
    \IEEEauthorrefmark{2}School of Information, Central University of Finance and Economics, China.\\
    \IEEEauthorrefmark{3}Singapore University of Technology and Design, Singapore.
    }
 }
\maketitle

\begin{abstract}
 Economic systems play pivotal roles in the metaverse. However, we have not yet found an overview that systematically introduces economic systems for the metaverse.
  Therefore, we review the state-of-the-art solutions, architectures, and systems related to economic systems.
  When investigating those state-of-the-art studies, we keep two questions in our mind: (1) what is the framework of economic systems in the context of the metaverse, and (2) what activities would economic systems engage in the metaverse? 
  This article aims to disclose insights into the economic systems that work for both the current and the future metaverse. 
  To have a clear overview of the economic-system framework, we mainly discuss the connections among three fundamental elements in the metaverse, i.e., digital creation, digital assets, and the digital trading market. After that, we elaborate on each topic of the proposed economic-system framework. Those topics include incentive mechanisms, monetary systems, digital wallets, decentralized finance (DeFi) activities, and cross-platform interoperability for the metaverse. 
  For each topic, we mainly discuss three questions: a) the rationale of this topic, b) why the metaverse needs this topic, and c) how this topic will evolve in the metaverse.
  Through this overview, we wish readers can better understand what economic systems the metaverse needs, and the insights behind the economic activities in the metaverse.

\end{abstract}


\section{Introduction}

 Metaverse has seized enormous attention from both academia and industries. Existing technologies that are enabling metaverse can be briefly classified into two categories, i.e., a) \textit{how to build the metaverse}, and b) \textit{how to enter into and how to experience the metaverse}. For the former category, technologies include digital twins, game, 3-dimensional (3D) rendering, artificial intelligence (AI) algorithms, etc. As for the latter one, the related technologies involve interactivity technologies such as virtual reality (VR), augmented reality (AR), mixed reality (MR),  human-computer interfaces, etc.

 In addition to those two categories, metaverse also requires support from the bottom-layer infrastructures \cite{yang2022fusing}, including networking and computing, dedicated operating systems, and blockchains. Firstly, networking and computational resources enable high-bandwidth and low-latency immersive experiences for metaverse users. Secondly, operating systems provide the system-level execution environment for the applications (Apps) of metaverse. Thirdly, blockchains can handle the high-volume transactions submitted by metaverse users when they are using metaverse Apps.

\begin{figure}[t]
        \centering     \includegraphics[width=1.\linewidth]{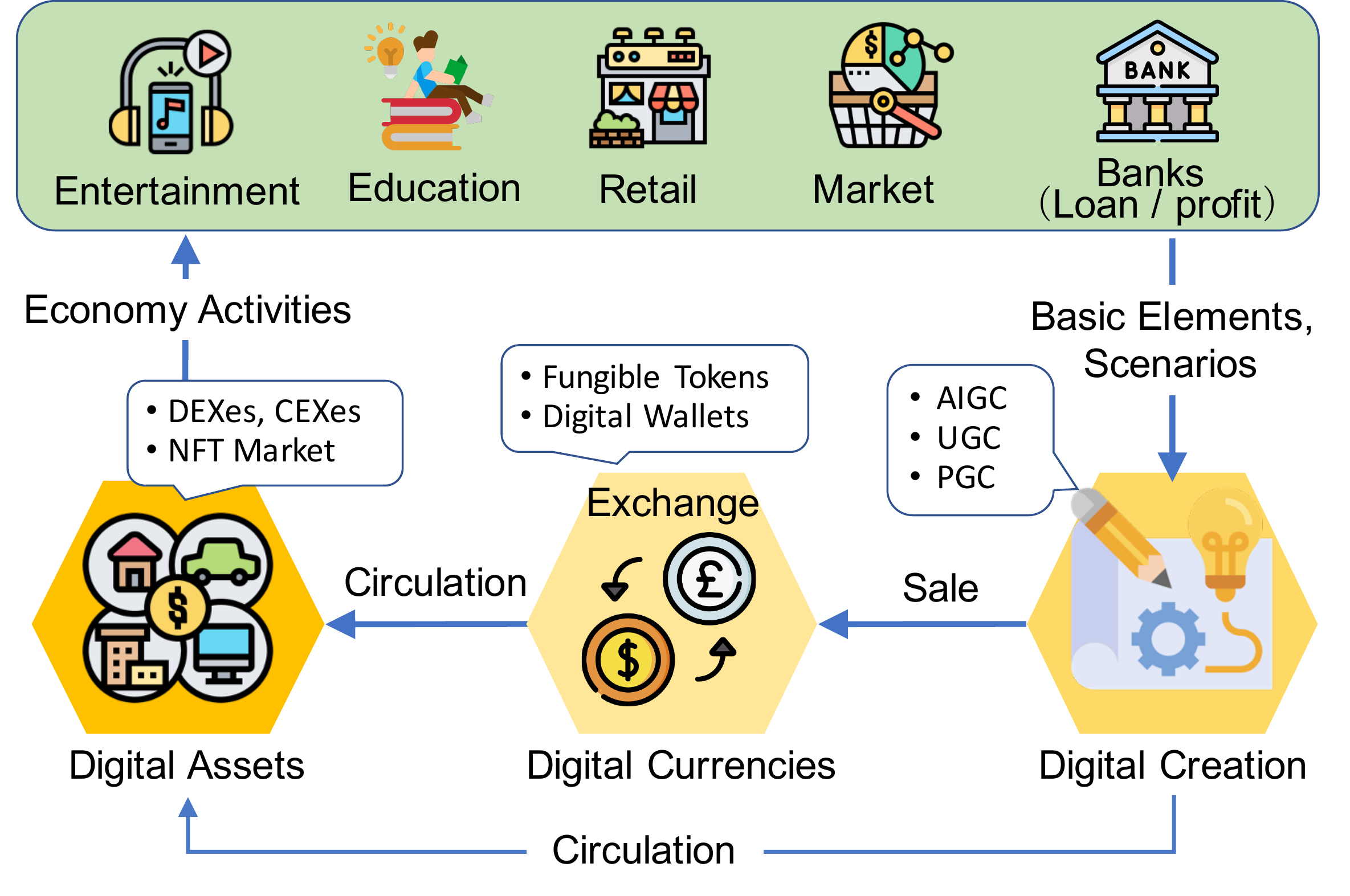}
        \caption{Framework of economic systems for the metaverse.}
        \label{fig:framework}
 \end{figure}

 The metaverse is not a virtual world that is independent of the real world, it is the extension of the physical world. Both worlds merge together to create an integrated ecosystem \cite{tlili2022metaverse}.
 In this article, we concentrate on blockchain-based economic systems in the context of the metaverse. As shown in Fig. \ref{fig:framework}, we present our overview of the economic systems that are promising to enable any possible economic activities and issues in both the current and future metaverse.

 Since Bitcoin \cite{nakamoto2008bitcoin}, blockchains have been exploited to implement economic systems for various sectors, including decentralized finance (DeFi), cryptocurrencies, non-fungible tokens (NFT), the trading platforms of digital assets, exchanges, etc.
 In the future, the metaverse will be a virtual living habitat for humans, all users create a complete supply chain to produce and consume digital content collaboratively. In order to protect the ownership of digital products, the owners need to control and track their circulation process. Blockchain, as the feasible decentralized infrastructure of Web3.0, enables users to trade digital assets with the property of transparency and traceability. What's more, smart contracts enable programmability and automated execution of transactions. By minting digital products into NFT, all stakeholders can control the ownership of digital products and share the economic value of the metaverse. Encouraged by sharing economics, a large number of digital products are generated by users in a decentralized approach. The associated economic activities imply that the blockchain will play a crucial role in the economic systems of the future metaverse.

 Although the economic systems existing in the current physical world have been deeply developed, we are wondering the question what  economic systems should look like in both the current and the future metaverse.
 The reason why we are curious about this question is that we think the economic systems are the foundation of metaverse. Almost every activity that occurs in metaverse is related to the economy. In the virtual world of metaverse, the economic systems need to handle economic activities that are severely different from those in the real world \cite{chen2022digital}. It indicates that the real-world economic systems will not be applicable to the virtual-world economic environment in metaverse.
 However, we have not yet found a survey article that systematically discusses such a topic. To fill this gap, we are motivated to conduct this overview of economic systems and wish to contribute some viewpoints, thoughts, and insights to the communities of the metaverse.

 The organization of this article is described as follows.
 Section \ref{sec:preliminaries} depicts the preliminaries of economic systems in the metaverse. 
 Section \ref{sec:MonetarySystem} explores the role of monetary systems for the metaverse.
 Section \ref{sec:IncentiveMechanisms} talks about the incentive mechanisms for the metaverse.
 Section \ref{sec:economicActivities} discusses the typical economic activities such as DeFi in the metaverse.
 Section \ref{sec:Cross-chain-Ecosystem} shows the insights of the cross-metaverse interoperability for the metaverse.
 Section \ref{sec:openIssues} discusses the challenges of open issues of economic systems in the metaverse.
 Finally, section \ref{sec:conclusion} concludes this article.

\section{Preliminaries of Economic Systems in Metaverse}\label{sec:preliminaries}


\subsection{Virtual-Reality Interactions for Metaverse}

\begin{figure}[t]
        \centering     \includegraphics[width=1.\linewidth]{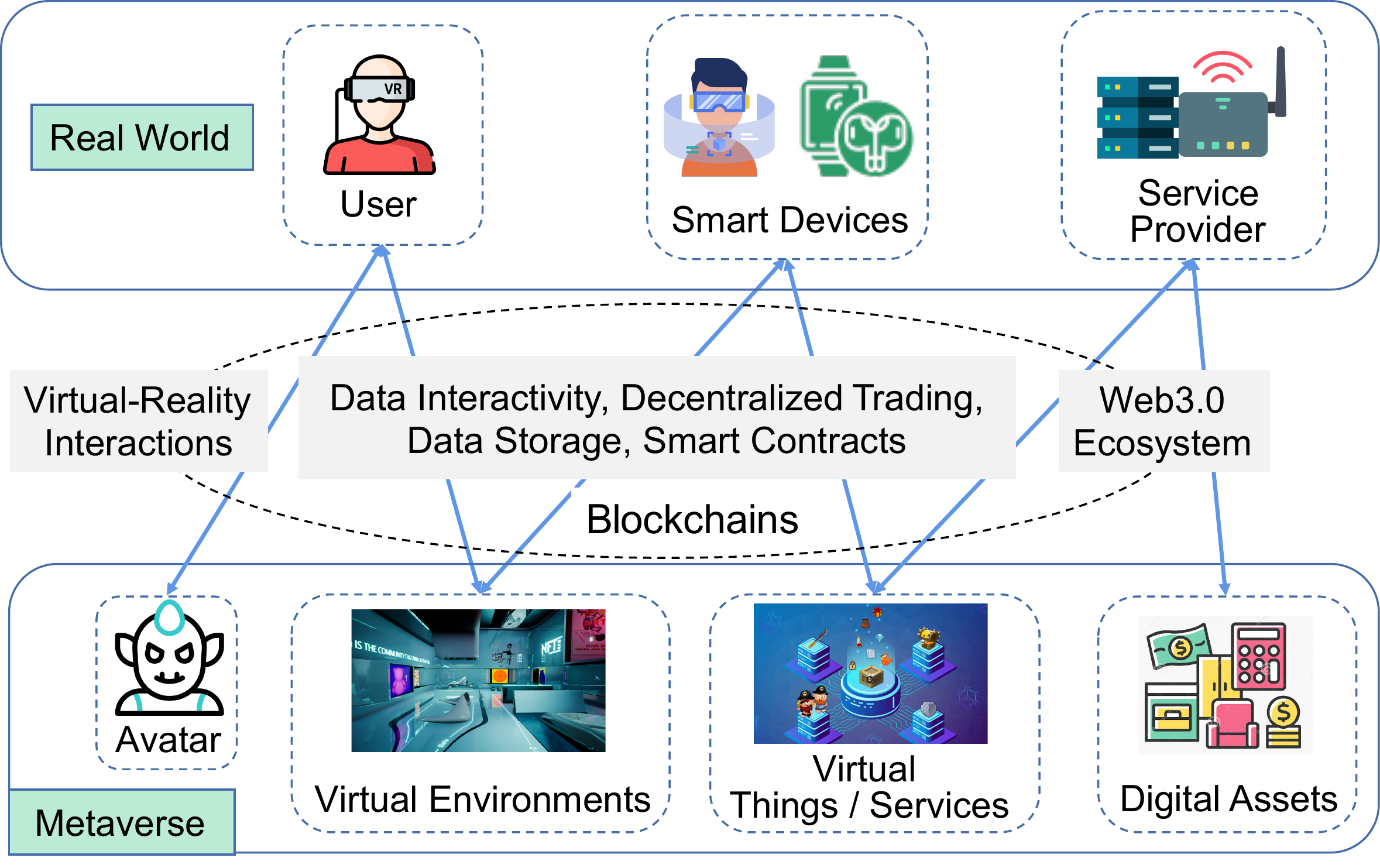}
        \caption{Virtual-reality interactions between physical world and metaverse.}
        \label{fig:vri}
 \end{figure}

 We first show an overview using Fig. \ref{fig:vri}, in which the virtual-reality interactions are demonstrated between the real-world objects and those in the metaverse. In the real-world layer, the objects include users, smart devices such as AR/VR/MR that help users enter into the metaverse, service providers, etc. In the metaverse layer, related objects include virtual things, virtual services, virtual environments, digital assets, etc. We then summarize all essential interactivity between those two layers as follows.
    \begin{itemize}
        \item Users and their avatars need blockchain-based services to have virtual-reality interactions.
        \item Users and their smart devices require data storage services, data interactivity applications, and trading systems, to interact with the virtual objects, virtual services, and virtual environments in the metaverse.
        \item Users manage their digital assets through the smart contract-enabled web3.0 ecosystem.
    \end{itemize}

 \subsection{Technology Companies Investing in Metaverse}
    
  Metaverse provides an interconnected virtual community that is changing social media and video game platforms. People interact in metaverse by using VR headphones, AR glasses, and the associated applications. Users can try on clothes, hang out in the virtual community, or even work in their virtual offices \cite{MetaverseBlockchain}.

  Some of the top technology companies are involved in the metaverse. Meta (previously Facebook) plans on developing an engaging and immersive social-interaction experience in the metaverse. Meta has launched a meeting software called Horizon Workrooms with the use of Oculus VR headsets. Nvidia designs and manufactures hardware for creating and supporting web3 applications. Nvidia's Omniverse helps web3 developers create projects in their metaverse \cite{CompaniesMetaverse}.
    
  Video game companies also are playing a leading role. Epic Games raised \$1 billion for the long-term construction of the metaverse. The game platform Roblox \cite{Roblox} describes its blueprint for the metaverse, in which people can learn, work, play, create and socialize together in millions of 3D experiences \cite{GameMetaverse}.



 \subsection{Digital Assets in the Metaverse}
 
 Van Niekerk \cite{van2006strategic} clearly defines the digital asset as ``any item of text or medium that has been formatted into a binary source that includes the right to use it". From this perspective, the concept of \textit{digital asset} was born not due to the driving power of internet technologies, but due to the power of the \textit{digital citizenry} concept coming to life \cite{toygar2013new}.
 
 Digital assets are the engine to drive the continued development of the economic system of the metaverse. The aim of this subsection is to introduce typical digital assets and the way how to create them in the metaverse.

    \subsubsection{User-Generated Content (UGC)}
    
    UGC is any type of digital content generated by metaverse users, including pictures, music, videos, etc. The generated content contains personal privacy data and potential economic value \cite{lee2021creators}.
    UGC is a promising alternative tool to identify the demand of users. It is mainly based on ordinary user-generated content, starting from users' needs, and everyone can publish content on the platform. 
    When the content is approved by the system or manually, it can be displayed on the platform to the audience.
    Similar to the concept of \textit{We Media}, users can create a variety of personal digital content, including blogs, podcasts, news, and videos. 
    In metaverse, UGC tends to be heterogeneous which triggers the surging demand for the ownership of UGC \cite{duan2021metaverse}.
    However, the quality of user-generated content varies since there is no requirement for the skills of creating.
    The existing methods are inefficient for a large number of UGCs because much digital content is highly informal and duplicated. 
    Timoshenko \textit{et al.} \cite{timoshenko2019identifying} apply a deep learning-based approach to filter out noninformative content to avoid sampling repetitive content. 
    
    Although some researchers have focused on related research questions for UGC in the metaverse, there are still some challenges, e.g., ownership control, payment scheme, and incentive mechanism.

 \subsubsection{Professional Generated Content (PGC)}
    
 PGC means the content is generated by experts or professional institutions that have professional content production capabilities. This manner can ensure the professionalism of the content.
 Therefore, PGC is generally checked by the platform. It is generally original content and pays more attention to copyrights. PGC can ensure the value and competitiveness of the content. For example, PGC video services is an user-friendly advertisement environment \cite{kim2012institutionalization} with the characteristics of specialization and commercialization. However, PGC services have some drawbacks, including geographical restrictions and a lack of professional user participation. The threshold for professional content creation is relatively high. That is why there are corresponding charges in some knowledge payment platforms. 
    
 For paid content, piracy is rampant. This phenomenon incurs a loss to both the platform and the payer of content. The platform needs a set of strict audit standards to ensure the quality of the content and must be able to produce high-quality content continuously. For PGC, platform procurement costs are higher than that of UGC.

  \subsubsection{Artificial Intelligence-Generated Content (AIGC)}
  
  With the breakthroughs of Artificial Intelligence (AI), natural language generation (NLG) technology can be applied to the digital content generated by the metaverse, such as news reports, poetry, and photo-generated. Nils \textit{et al.}  \cite{kobis2021artificial} adopt the NLG algorithm GPT-2 to generate poem samples by identifying the character of human poems.
  AIGC is a typical manner to produce digital assets \cite{wang2022survey}. In the future, with the development of the metaverse, the number of digital content consumers will far exceed that of digital content producers.
   
   The AIGC enables metaverse to create massively qualified and customized content. Generally, AIGC consists of content creation in two ways \cite{wang2022survey}, i.e., {\itshape 1)} AI generates digital content independently, and {\itshape 2)} users with assistant AI create digital content. 
 For example, Epic Games create a large number of virtual roles (e.g., virtual conversational assistants) by AI algorithms in \textit{MetaHuman} \cite{lyytinen2021metahuman}. Singer \textit{et al.} \cite{singer2022make} propose an approach named \textit{Make-A-Video}, for directly translating the tremendous recent progress in Text-to-Image (T2I) generation to Text-to-Video (T2V). With Make-A-Video, the generated videos inherit the vastness (i.e., diversity in aesthetic, fantastical depictions, etc.) of today’s image generation models. However, Make-A-Video currently can only generate 5 seconds of 16 frames per second silent clip. The picture can only describe one action or scene, and the pixels are only 768$\times$768. Meanwhile, Ho \textit{et al.} \cite{Ho2022ImagenVH} present a text-conditional video generation system named \textit{Imagen Video}, which is based on a cascade of video diffusion models. Given a text prompt, Imagen Video generates high-definition videos using a base video generation model and a sequence of interleaved spatial and temporal video super-resolution models. Feng \textit{et al.} \cite{Feng2022ERNIEViLG2I} propose a large-scale Chinese text-to-image diffusion model, named \textit{ERNIE-ViLG 2.0}, which progressively upgrades the quality of generated images by i) incorporating fine-grained textual and visual knowledge of key elements in the scene, and ii) utilizing different denoising experts at different stages.
    Dong \textit{et al.} \cite{Dong2022DreamArtistTC} propose a method, called \textit{DreamArtist}, which employs a learning strategy of contrastive prompt-tuning. DreamArtist introduces both positive and negative embeddings as pseudo-words and trains them jointly. With DreamArtist, everyone can be an artist who has productive imagination, specialized experiences, and fantastic inspirations.
    Wu \textit{et al.} \cite{wu2020investigating} investigate the explicit and implicit perceptions of AI-generated poetry and painting held by subjects from two societies (The USA and China).

    It can be found that the aforementioned works focus on AI-generated content with respect to photos, videos, text, etc. While existing AI products are still far from human creation in terms of visuals and storylines, Meta and Google's new products are really impressive and raise the question of how AI will lead content production. However, skeptics and proponents continue to argue whether AI-generated content will ultimately satisfy the benchmark of content produced by human writers \cite{latar2015robot}.

\begin{figure*}[t]
     \centering
     \begin{subfigure}[b]{0.49 \textwidth}
         \centering \includegraphics[width=0.97\textwidth]{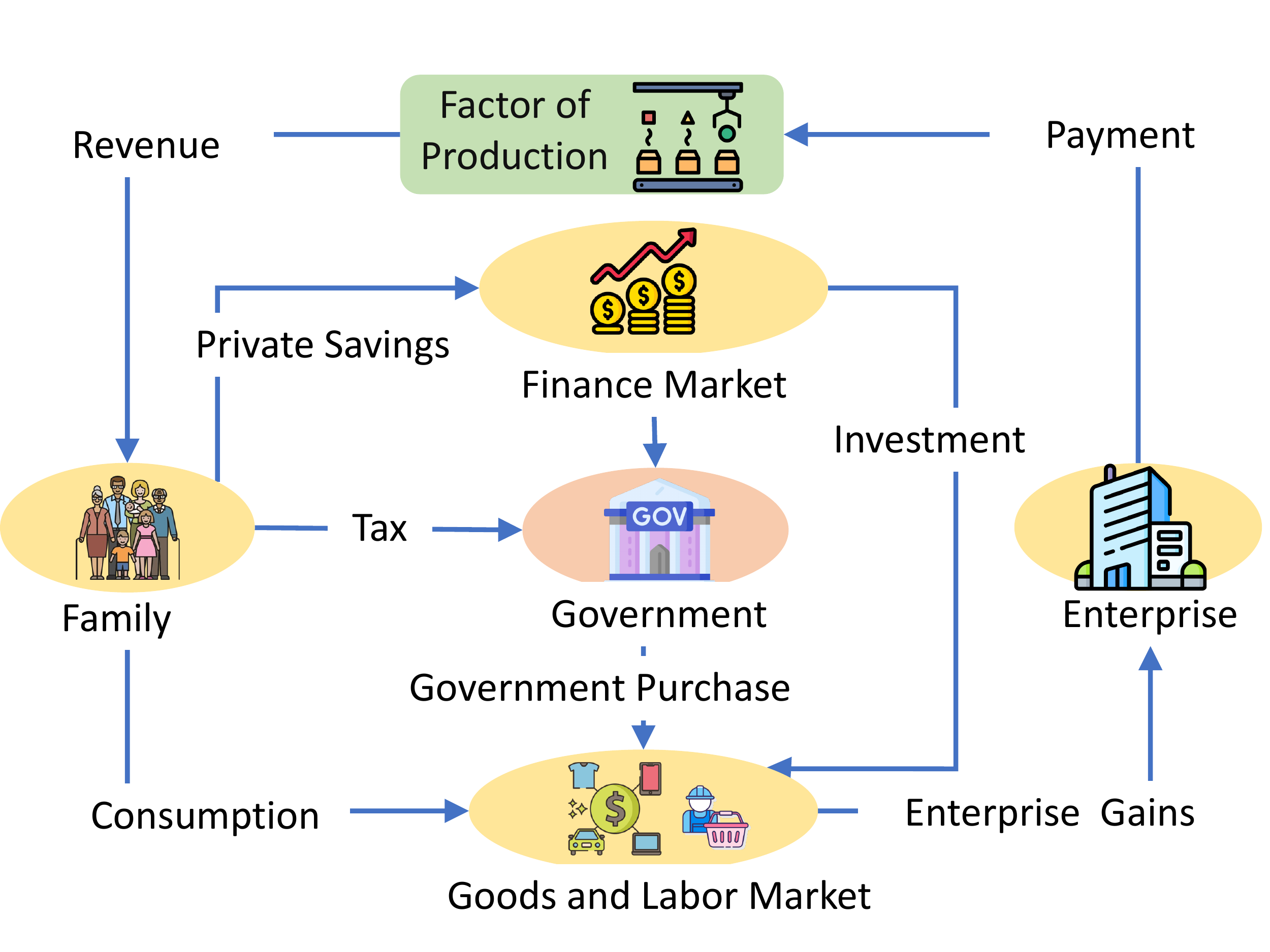}
    \caption{Macroeconomic circulation in physical world}
    \label{fig:currencyCycling}
     \end{subfigure}
     \hfill
     \begin{subfigure}[b]{0.49 \textwidth}
         \centering
        \includegraphics[width=0.95\textwidth]{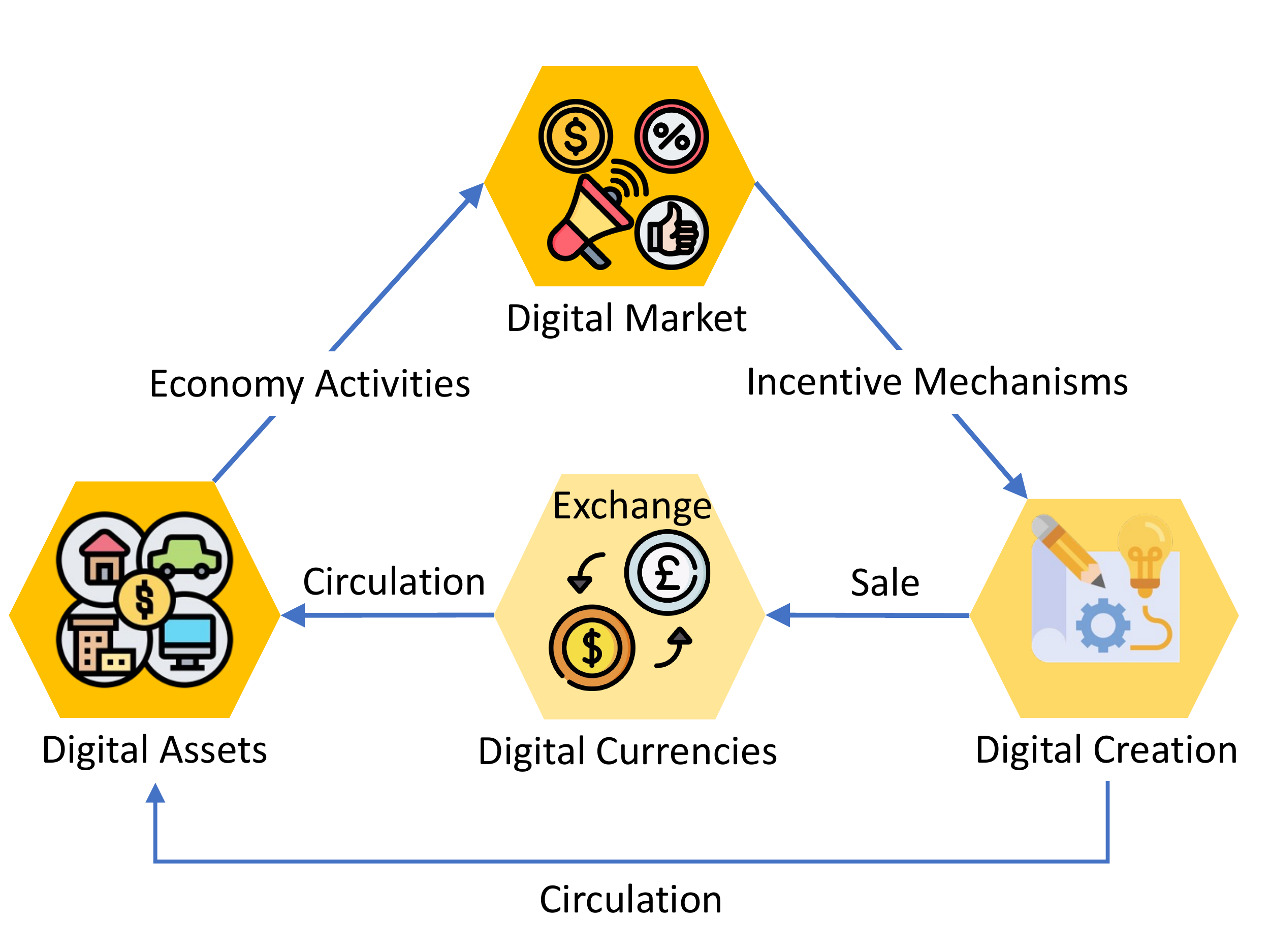}
    \caption{Currency circulation in metaverse}
    \label{fig:metaEconomic}
     \end{subfigure}
     \caption{Comparison of economic systems between the physical world and the metaverse.}
        \label{fig:Comparisonof economicsystem}
\end{figure*}

 \subsubsection{Non-Fungible Tokens (NFT)}
  
     NFT is a typical category of digital assets based on blockchains. NFTs ensure the uniqueness of digital assets by permanently storing historical encrypted transactions on the associated blockchain \cite{xu2022full}.

     As a blockchain-based digital token as a proof of ownership and authenticity for crypto assets, NFT offers a futuristic possibility for art trading \cite{lin2022dai}. For example, Minecraft \cite{niemeyer2015maker} has a complicated economic system that allows players to accumulate online Axie Infinity \cite{axieinfinity}, which is an NFT-driven game built on Ethereum.
     NFT can tokenize multiple types of digital assets like art, music, collectibles, video, and in-game items to guarantee uniqueness and authentication \cite{wang2021non}.
     NFT plays an important role in determining authentic rights for metaverse assets \cite{lim2022realizing}. Specifically, users can store their digital assets as NFT on blockchains and trade digital assets through smart contracts.

    NFTs have strong cultural and interactive attributes. If a user purchases an NFT, he/she will be the only owner in the world to hold and prove that asset. Such a purchase behavior has strong social significance, by which consumers can demonstrate their unique purchasing ability, taste, and even social status in the digital field. 
    Therefore, the recognition of the value and ownership of NFT on the basis of the technical blockchain requires a rich social activity and the consensus of a certain number of participants. NFT has everything to do with digital collections that realize the asset of virtual productions so that digital assets have tradeable entities.

    The cost of minting an NFT includes the gas fee, account fee, and service fee. The gas fee is used to pay for dealing with the transaction on blockchains. Account fee means a small fee that a platform charges users to place their products on its platform.
    In basic finance, service fee denotes the fee that auctioneers, salespeople, and others are paid a commission for their services. A flat commission is charged to the seller.

\subsection{Trading and Market of Digital Assets in Metaverse}

    The digital market includes all exchange activities of products and services that rely on Internet-based digital technologies. This market is composed of digitizing traditional products and services, such as e-commerce and online marketplaces that simply move offline transactions directly online. At the micro level of the economic system for the metaverse, clients are the foundation of economic activities, working as both producers and consumers of user-generated content \cite{chen2022economics}.

 Meanwhile, as depicted in Fig.\ref{fig:Comparisonof economicsystem}, the exchange is an intermediary bridging production and consumption. 
 Compared with Fig.\ref{fig:currencyCycling}, we can find that the metaverse economy is powered by blockchain and cryptocurrency technologies. As shown in Fig. \ref{fig:metaEconomic}, this new fashion of economy differs from the traditional financial system. The decentralized metaverse economy provides financial products to users without involving intermediaries such as banks, brokerages, or insurance companies.
    
 For trading digital assets, Hasan \textit{et al.} \cite{8501910} emphasize that proof of delivery (PoD) of the digital content is an immense need since these assets are subject to payment. 
 To support the requirement of immutable and tamper-proof logs, accountability, and traceability, the authors propose a decentralized PoD by leveraging major features of blockchain and Ethereum smart contracts. Blockchains before Ethereum, such as Bitcoin, only support token transferring. Until the emergence of Ethereum, smart contracts begin to support Turing-complete programming. Complicated businesses could be executed in a virtual machine through smart contract codes. 
 Ethereum realizes the upgrade of blockchain applications from cryptocurrency to crypto business.

 Based on smart contracts and crypto tokens, decentralized finance (DeFi)  offers a new approach to innovate economic models in the metaverse. Empowered by such advanced blockchain technologies, DeFi can boost the decentralized market and business in the metaverse. Existing successful solutions, such as Uniswap \cite{angeris2019analysis}, a decentralized exchange (DEX) implemented on Ethereum, automatically provide users with liquidity for their metaverse tokens. 
 We review a representative study related to the DeFi market and business here. More discussion can be found in Section \ref{sec:economicActivities}.
 In \cite{daian2020flash}, the authors analyze the behavior of arbitrage bots in the context of the cryptocurrency market. They find that arbitrage robots could observe the transactions in the transaction pool and perform arbitrage without risks. They also present a cooperative strategy to maximize the profit of arbitrage robots and point out that miners could act as arbitrage robots under certain circumstances. However, the miner extractable value (MEV) could incentivize the emergence of forking attacks. The authors propose a cooperative bidding strategy for them to strive for more profit. They also find that the current amount of MEV in a month is more than 25$\times$ the cost of a 51\% attack on Ethereum.

\subsection{Metaverse and Blockchains}

 \subsubsection{An Open Metaverse Needs Blockchains}
 
    Metaverse is a decentralized virtual world based on blockchains. This means that the future metaverse will not be managed by a centralized entity, but is maintained by a large number of decentralized entities around the world. Therefore, metaverse technology is not subject to government review and is convenient for user experience \cite{MetaverseDefine}. 

    The metaverse can exist without both blockchains and cryptocurrencies. However, an open and rational metaverse should be linked to open blockchains with highly interoperability, in which virtual assets can be exchanged and circulated in a trustless and decentralized manner.

  Blockchains are used to record any type of data in publicly distributed ledgers \cite{zhou2020solutions}. From such ledgers, people can track the historical transactions of cryptocurrencies, NFTs, and other digital assets recorded in the chain. Given that the centralized token systems have suffered from severe hacking threats, centralized cryptocurrency exchanges (CEXes) may be stolen \cite{FaceBookMetaverse}. Therefore, the blockchain-based token system is more reliable than CEXes.
 
\subsubsection{High-Performance Blockchains Are the Foundation of Future Metaverse}

 Considering that the scalability of the metaverse keeps growing exponentially, it is not hard to imagine that the transactions originating from the future metaverse will be much more intensive than those in real-world scenarios. Thus, the future metaverse needs high-throughput and highly reliable blockchains, aiming to handle a giant number of transactions.

 A lot of high-throughput blockchains have been proposed, such as Solana \cite{solana}, Avalanche \cite{Avalanche}, Linera \cite{Linera}, Sui \cite{Sui}, Aptos \cite{Aptos}, and BrokerChain \cite{huang2022brokerchain}. Those blockchains claim to provide faster speeds, higher throughput, and lower fees than conventional blockchains such as Ethereum. Some of those new blockchains are also compatible with Ethereum assets and dApps. Given those advantages, we still believe that today's high-throughput blockchains will not be applicable to the requirement of the future metaverse. Therefore, researchers are encouraged to design new powerful blockchains that can achieve super high throughput such as at least 100 thousand \textit{transactions per second} (TPS), and super high reliability.


\section{Incentive Mechanisms for Metaverse}\label{sec:IncentiveMechanisms}

    \subsection{The Role of Incentive Mechanism in Metaverse}
    
\begin{figure*}[t]
        \centering     \includegraphics[width=0.9\linewidth]{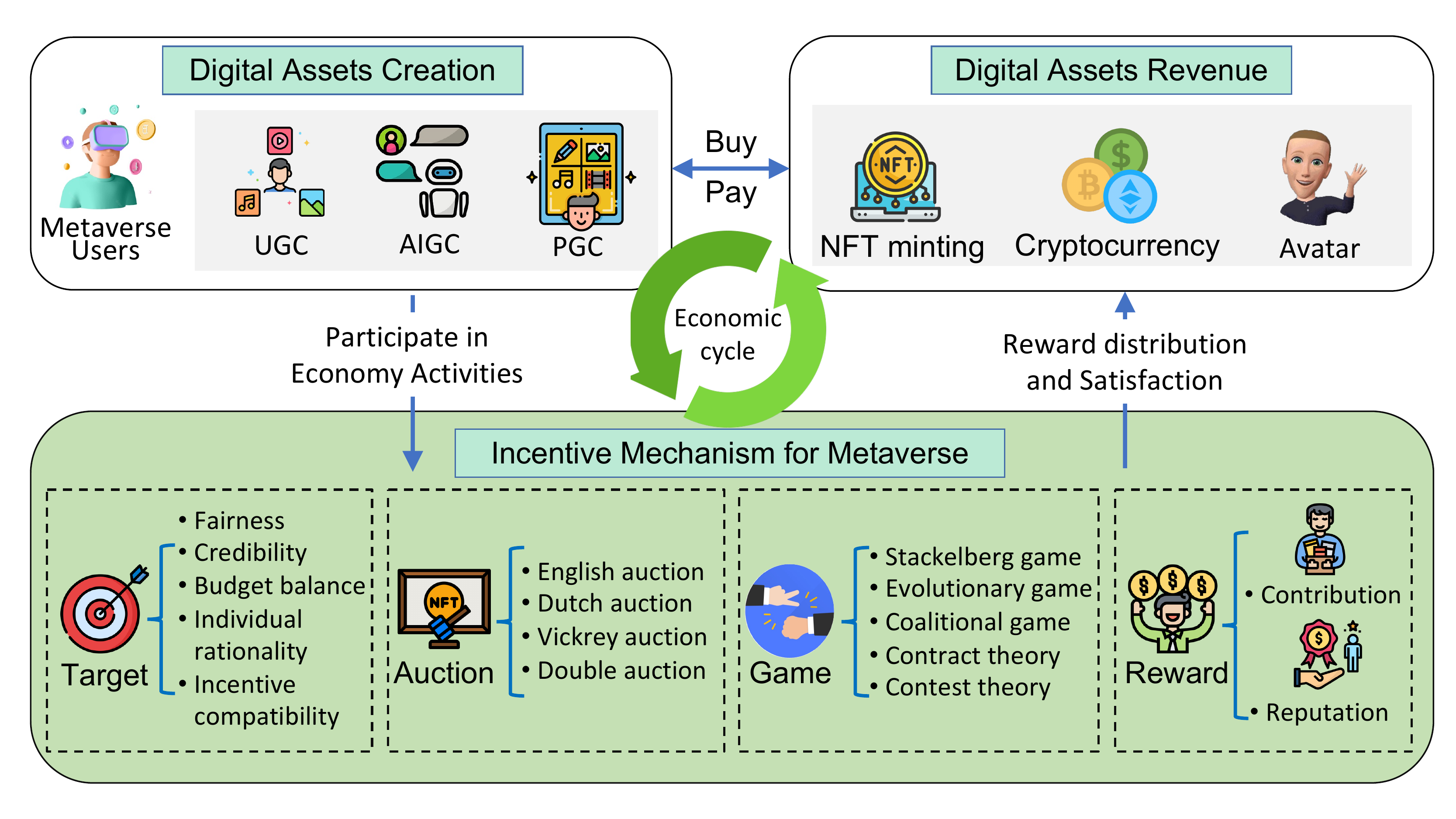}
        \caption{Incentive mechanisms for metaverse.}
        \label{fig: incentive mechanism}
    \end{figure*}

    The economic system established by metaverse users has become a promising topic for driving innovations toward the metaverse. Users, as essential stakeholders, contribute resources and data to create a virtual world, where users own a large number of digital assets. In this regard, incentive mechanisms need to be designed to subsidize the contribution of users and encourage all stakeholders to benefit from the metaverse economic system. The economic cycle of digital assets is supported by incentive mechanisms as shown in Fig. \ref{fig: incentive mechanism}. Metaverse users can generate a great number of digital content (i.e., UGC, AIGC, PGC) independently or with AI assistance. Users participate in economic activities  (e.g., creation, exchange, and investment) to get some revenue. The incentive mechanism is responsible for the reward allocation fairly to support the metaverse economic system.

    Without a series of reasonable incentive mechanisms, it is obvious that users are unwilling to contribute computation and data resources to participate in metaverse service computation tasks under the risk of privacy disclosure.

    \subsection{Design Target  of Incentive Mechanism}

    The design target of incentive mechanisms is to motivate more high-quality contributions through various forms of extrinsic rewards to meet the personal needs of contributors. Therefore, in order to make the incentive mechanism work continuously and effectively, the following properties should be considered. 
    \begin{itemize}
        \item \textit{Fairness}: Fairness may dramatically affect the optimal results of incentive mechanism \cite{fehr2000fairness}. The fairness of reward distribution determines the sustainability of the incentive mechanism. Han \textit{et al.} \cite{yu2020fairness} design a fairness-aware incentive based on the interest of data owners, which provides three fairness criteria, including contribution, regret distribution, and expectation. Existing works focus on the fairness of incentive mechanisms to prevent free-rider behavior, which refers to participants trying to earn income, but no contribution \cite{gao2015survey, zhu2016fair, sinha2017incentive, li2019considering, zhu2021impact}.
        \item \textit{Credibility}: The credibility of the optimal strategy directly affects the execution effect of the incentive mechanism. Mart{\'\i}n-Herr{\'a}n \textit{et al.} \cite{martin2005credibility} characterize the credibility of incentive equilibrium strategies with linear-state games. BESIFL \cite{xu2021besifl} is a blockchain-empowered decentralized federated learning paradigm. Blockchain is adopted to achieve malicious node detection and incentive management. Some existing truthful incentive mechanisms are designed to motivate users to devote resources to collaborative computation and offloading \cite{zhang2015truthful, he2019truthful, wang2017towards, liwang2018truthful}.
        \item \textit{Budget Balance}: The incentive mechanism satisfies budget balance if and only if the payment of buyers is non-negative and the whole system does not need investments from another third party \cite{ott1965budget}. In other words, the summation of the money transfer between all parties is zero \cite{tang2021incentive}. An auction satisfies the budget balance if and only if the payment collected from all buyers is at least as large as the payment to the sellers \cite{yang2013truthful}.
        \item \textit{Individual Rationality (IR)}: The incentive mechanism satisfies IR if and only if the revenues of all participants are non-negative \cite{gode1993allocative}. In the metaverse, rational resource pricing can improve the quality of the immersive experience of users, where users purchase bandwidth resources to reduce the communication latency of services, and metaverse service providers decide the price of the resource to achieve individual rationality.
        \item \textit{Incentive Compatibility (IC)}: The incentive mechanism that satisfies IC is expressed as $\gamma$-truthfulness \cite{myerson1979incentive}. The reward information submitted by the platform is based on the evaluation of contributors' task value, while the contribution information submitted by  contributors is calculated by their resource consumption. It has been reviewed that Bayesian incentive compatibility, efficiency, and budget balance of the standard Arrow-d'Aspremont-Gerard-Varet (AGV) mechanism \cite{wang2016fair}. Moreover, Ma \textit{et al.} \cite{ma2014incentive} propose an enhanced AGV mechanism to achieve IC, IR, and budget balance.

    \end{itemize}

    Most of the existing incentive mechanisms toward the metaverse ecosystem are designed based on economic theories, including auction mechanisms, game theory-based strategy optimization, reward systems, and reputation mechanisms. 
    The existing representative incentive solutions are reviewed in the following subsections.

    \subsection{Auction Mechanism}
    
    Auction has been regarded as a promising solution \cite{jin2015auction} to design incentive mechanisms. In the metaverse, auction mechanisms can assist buyers in bidding on the valuation of digital content to achieve an efficient circulation of digital assets. However, the heterogeneity of digital assets and the potential risk of privacy disclosure may lead to an unsustainable auction market and some issues, such as the winner's curse \cite{thaler1988anomalies} and unfair bids \cite{loosemore2015inter}. There are some existing researches that focus on designing more efficient auction mechanisms to motivate users to participate in metaverse economic activities. The existing auction mechanism can be classified into the following categories as described in Table \ref{Table:auction}

\begin{table}[t]
\caption{Auction Mechanisms and Representative Examples}
\centering
\footnotesize
\begin{tabular}{|m{1.9cm}<{\centering}|m{3.3cm}|m{2.5cm}|}%
\hline
\textbf{Auction Name} & \textbf{Description} & \textbf{Examples} \\
\hline 
    English Auction & The classical public sale method, bidding starts from the lowest price. & Jiao \emph{et al.} \cite{Jiao2018Welfare}, Zhang \textit{et al.}\cite{zhang2021privacy}.\\
    \hline
    Dutch Auction & Reversing to the English auction, bidding starts from the highest price. & Fan \emph{et al.} \cite{fan2020hybrid}, Xu \emph{et al.} \cite{xu2021wireless}. \\
     \hline
     Vickrey Auction& Also known as the sealed second-price auction, the final price is the second-highest bid. & EPViSA \cite{xu2022epvisa}, Zhang \emph{et al.} \cite{zhang2022truthful}, Luong \textit{et al.} \cite{Luong2018Auction}.\\
     \hline
     Double Auction & Multiple buyers and sellers offer bids, and the final price is decided by matching the bids in a certain order. & Ng \emph{et al.} \cite{ng2021double}, Kim \emph{et al.} \cite{kim2022auction}, Wang et al. \cite{wang2019privacy}, Liew \textit{et al.} \cite{liew2022economics}.\\
     \hline
\end{tabular}
\label{Table:auction}
\end{table}
    
    In metaverse services, the seamless, immersive, and interactive user experience needs to rely on real-time data synchronization between multiple physical entities. Jiao \textit{et al.} \cite{Jiao2018Welfare} propose a classical auction-based resource allocation mechanism for the edge computing service provider to maximize social welfare while achieving credibility, computational efficiency, and individual rationality. Fan \textit{et al.} \cite{fan2020hybrid} develop a resource trading platform based on a hybrid blockchain, in which a reverse auction mechanism is executed automatically by a blockchain-based smart contract. In order to further improve the virtual driving experience, it's essential to design an effective allocation mechanism for synchronizing real-time data from autonomous vehicles to roadside units (RSUs). To this end, EPViSA \cite{xu2022epvisa} is an enhanced second-score auction-based data synchronization scheme where the physical and virtual entities can synchronize data and resource in the vehicle-metaverse service market. The trading rule of the second-score auction belongs to a two-dimensional analog of the Vickrey auction. When utilizing digital twins to construct a virtual, there remains limited research on how to allocate station resources. Zhang \emph{et al.} \cite{zhang2022truthful} solve the problem of the optimal resource allocation in the wireless channel. payment rule based on the Vickrey Clarke Groves (VCG) auction is adopted to decide the payment rule of resources to maximize social welfare. Moreover, existing researches show that the double auction in resource allocation can achieve preferable economic properties. Ng \textit{et al.} \cite{ng2021double} adopt the double auction to assist in the resource allocation of the edge servers and determines the prices of resources to complete the Coded Distributed Computing tasks.

    Deep learning has proven effective in optimizing the auction mechanism. Luong \textit{et al.} \cite{Luong2018Auction} optimize the auction for edge resource allocation by a multi-layer neural network, which confirms the advantage that deriving high revenue auction using deep learning. Xu \textit{et al.} \cite{xu2021wireless} propose a double dutch auction-based  incentive mechanism for VR service of the metaverse, in which deep reinforcement learning (DRL) is adopted to determine optimal pricing strategies and allocation rules of VR services. Kim \textit{et al.} \cite{kim2022auction} formulate the rule of data resource trading between IoT service providers and  edge devices by McAfee double auction, in which a Q-learning algorithm is leveraged to evaluate the power level of edge devices. Liew \textit{et al.} \cite{liew2022economics} adopt deep learning and double auction to address the energy allocation for IoT devices, in which the revenue of semantic service providers can be maximized. Through the combination of auction and learning algorithms, the interaction of system agents can be modeled and optimal to achieve the balance of the metaverse market.

    Some researchers have paid attention to the privacy-disclosure problem in the auction mechanism. For example, Zhang \textit{et al.}\cite{zhang2021privacy} propose a privacy-preserving auction mechanism for data aggregation in mobile crowdsensing tasks, where the platform as auctioneer recruits workers to complete sensing tasks. Wang \textit{et al.} \cite{wang2019privacy} proposed a privacy-preserving and truthful double-auction mechanism named PS-TAHES based on additive homomorphic encryption \cite{paillier1999public}, aiming to prevent personal privacy-information leakage in the auction. In general, research on auction mechanisms in the metaverse is still in its infancy and has not paid sufficient attention to privacy-preserving concerns.

    \subsection{Game Theory-based Strategy Optimization}

    Game theory has been widely adopted in the existing incentive mechanisms. In the economic system of the metaverse, game theory can model the interaction among rational agents participating in the metaverse economic activities and achieve utility and social welfare maximization by individual strategy optimization. Game theory can model interactive decision-making, where each player's strategy depends on the actions of all other players. A comparison of different game theory-based strategy optimization methods is presented as follows.

    Stackelberg game can construct a strategic model among leaders and followers, in which the leader move first, and then followers take action sequentially. In the vehicular metaverse, Jiang \textit{et al.} \cite{jiang2022reliable} construct a game-theoretic hierarchical architecture, in which the Stackelberg game is considered to motivate more reliable workers to participate in the intensive rendering computation. Liu \textit{et al.} \cite{liu2022incentive} formulate a three-stage Stackelberg game to motivate users to pay sufficient transaction fees, which mitigates the issue of insufficient blockchain revenue. Huang \textit{et al.} \cite{huang2022joint} model a Stackelberg game to price the resource service dynamically, in which metaverse service providers are leaders, and users are followers. The distributed and centralized approach is adopted to derive the Stackelberg equilibrium according to individual privacy requirements. Sun \textit{et al.} \cite{sun2021dynamic} construct a two-stage Stackelberg game to motivate users to participate in aerial-assisted Internet of Vehicles (IoV), which promotes the development of dynamic digital twins.  
    Jiang \textit{et al.} \cite{jiang2021reliable} adopt the coalition game and Stackelberg game to assist in choosing reliable workers to participate in the Coded Distributed Computing tasks in the metaverse.  Daniel \textit{et al.} \cite{daniel2022ipfs} adopt a two-stage Stackelberg game to analyze a Nash equilibrium with negative externalities and unfair prices for blockchain data storage.

    Evolutionary game is a typical dynamic game theory, which can model the dynamic decision process of evolution of individuals with biological characteristics in the metaverse. In the infrastructure layer of edge-intelligence empowered metaverse, Lim \textit{et al.} \cite{lim2022realizing} leverage evolutionary game to motivate the convergence of edge intelligence to support the metaverse engine. Specifically, the evolutionary game is used to model how the rewards from virtual service providers affect the contribution of sensing service providers during the service of the metaverse. The simulation shows that the synchronization frequency for virtual devices varies with the rewards.
    
    Coalitional game refers to the players participating in the game in the form of alliances and cooperation, which aim to identify the best coalitions and a fair reward distribution among all participants \cite{shams2013basics}. In economics, public goods refer to a commodity or service that is made available to many customers, which is both non-excludable and non-rivalrous. For public goods, like digital assets, the free-rider behaviors seriously hinder the enthusiasm of participants and fairness collaboration. Luo \textit{et al.}  \cite{luo2020incentive} design an efficient incentive mechanism to motivate edge devices to execute collaborative fog computing tasks and achieve mutually-beneficial resource cooperation based on the coalitional game theory. Considering the free-riding behaviors in collaboration, Pu \textit{et al.} \cite{pu2016d2d} adopt the coalitional game to formulate a time-average energy consumption minimizing task under incentive constraints to motivate more contribution in collaboration.
    
    Contract theory is another common method for modeling the interaction among players in the metaverse. Jiang \textit{et al.} \cite{kang2022blockchain} propose an age of information-based \textit{contract model} to motivate data sensing among industrial IoT devices for industrial metaverses, in which the age of information is introduced as the data freshness metric. Du \textit{et al.} \cite{du2022attention} provide an optimal contract design framework to model the interaction between the service providers and the network infrastructure providers in the metaverse, in which a novel metric named Meta-Immersion is proposed to  measure the subjective feelings of metaverse users. In the collaborative communication of end-edge-cloud, a blockchain-based incentive mechanism named InFEDge \cite{wang2022infedge} is designed to trade off the training overhead and model performance in the Hierarchical Federated Learning (HFL) based on contract theory. Taking the incomplete information into account, the authors obtain the optimal solution by formulating a contract game. The blockchain is leveraged to achieve reliable economic incentives and prevent disturbance from unreliable nodes. For vehicular edge computing, Liu \textit{et al.} \cite{liu2021blockchain} propose a smart contract-based incentive algorithm to encourage edge devices to contribute computation resources and evaluate the performance of Federated Learning.
    
    Contest theory is a special type of game where contestants exert costly effort to obtain some expected rewards with some probability, which is a very common phenomenon in the metaverse. To further improve metaverse service quality, Wang \textit{et al.} \cite{wang2022semantic} introduce \textit{contest theory} to create an incentive mechanism that motivates users to upload data more frequently. The authors design a semantic communication framework for sensing information from the physical world to the metaverse. 
    
    For the resource allocation of the metaverse, game theory can be well applied in the design of incentive mechanisms. Reasonable resource allocation supports seamless real-time immersive experience, especially during the data synchronization between physical entities and virtual worlds \cite{shen2021holistic}. Han \textit{et al.} \cite{han2022dynamic} adopt a dynamic evolutionary game to motivate different device owners to update resource allocation strategies, in which revenue maximization can be achieved by trading off the resource supplies and demands. Lin \textit{et al.} \cite{lin2021stochastic} design an effective congestion control incentive scheme for Digital Twin Edge Networks (DTENs), in which the long-term control decisions are decomposed into a series of online related decisions using Lyapunov optimization theory. In \cite{liu2020economics}, the authors propose a resilient incentive mechanism to trade off the storage requirements and mining revenues in the blockchain. Shen \textit{et al.} \cite{shen2019equity} propose a fair node resource allocation strategy to promote user cooperation and maintain reliable content storage collectively. Lin \textit{et al.} \cite{lin2019making} proposes a knowledge pricing strategy based on a non-cooperative game for the blockchain-based knowledge market, so that knowledge generated by edge nodes can be traded in  the edge-AI enabled Internet of Things. Hou \textit{et al.} \cite{hou2021incentive} formulate an incentive-driven resource allocation scheme to stimulate collaborative computing between edge servers and IIoT devices, in which an NP-hard problem is derived to optimize task assignment strategies and achieve utility maximization.

    \subsection{Fair and Credible Reward System}

    In a metaverse's economic system, the reward is one of the important ways to achieve incentive goals. Digital assets (i.e., cryptocurrencies, NFTs, and other tokens) are the main medium of rewards. These assets can be donated to the developers for the metaverse as an incentive mechanism \cite{huynh2023artificial}. Thomason \textit{et al.} \cite{thomason2022metaverse}propose tokenized incentives mechanism to motivate developers to participate in collaboration under micro-economic systems, aiming to make individual contributions consistent with the collaboration goals. For example, GameFi \cite{gamefi} refers to play-to-earn blockchain games that offer a potential economic incentive mechanism to players, where avatars controlled by players can earn cryptocurrency and NFT as the reward for accomplishing game tasks. In Sandbox \cite{Sandbox}, users can seamlessly access the virtual world, in which users vote for governance decisions via the decentralized autonomous organization (DAO) to earn token revenues.
    
    Reputation value is the main metric to evaluate the reliability of workers, and it can be used as the bases for the fair distribution of rewards. Jiang \textit{et al.} \cite{jiang2022reliable} formulate a reputation evaluation metric based on the subjective logic model to select reliable employees in distributed computing. The reputation value is stored in the blockchain database to ensure reliable management. In order to expand the methods of employee selection, Zhao \textit{et al.} \cite{zhao2020privacy} proposes a privacy-preserving incentive mechanism to encourage more IoT devices to participate in model training tasks and reduce malicious parameter updates. CoopEdge \cite{yuan2021coopedge} is proposed to solve the credible incentive mechanism of edge nodes based on blockchain, in which reputation management motivates selfish edge nodes to perform distributed edge computing collaboratively. In this problem, the historical performance of edge nodes in performing offloading tasks to evaluate the reputation, which can be recorded in the blockchain. The reputation value is calculated by a novel consensus mechanism named Proof of Edge Reputation (PoER), which allows miners to record reputation value on the blockchain by consensus validation.

    In addition, contribution is also viewed as one of the bases for reward allocation. Metachain \cite{nguyen2022metachain} is a decentralized metaverse framework based on blockchains, in which an incentive mechanism based on the Stackelberg game is constructed to attract more contributions during participating in the autonomous metaverse activities. Sun \textit{et al.} \cite{sun2021dynamic} propose a distributed incentive mechanism based on the alternating direction method to encourage more vehicles to contribute more resources to the physical-virtual synchronization in the metaverse. Their proposed mechanism can maximize the overall energy efficiency of vehicular in dynamic IoV scenarios.

\section{Monetary Systems for Metaverse}\label{sec:MonetarySystem}

\subsection{Basics of Cryptocurrency in Metaverse}


    Cryptocurrencies are the best-performed blockchain applications in the past few years. The term \textit{crypto} refers to encryption algorithms and techniques used to protect the native tokens of a specific blockchain system. The popular encryption methods exploited in blockchains include elliptical curve encryption, hashing functions, and public-private key signature technologies.

    Although the cryptocurrency market attracts enormous criticisms, people still see the value of cryptocurrencies in the ecosystem of blockchains, because cryptocurrency is a medium of value exchange in the digital world. Thus, cryptocurrencies can bring liquidity to the economic market of digital assets.

   
   
  Metaverse monetary systems require decentralized exchanges (DEXes) to enable the transactions of both UGC/AIGC and NFT. Currently, major economic activities in metaverse mainly involve the auction of virtual assets such as land, scarce items, and precious real estate, the development and leasing of land, rewards for accomplishing game tasks, and profits from investing in cryptocurrencies. Those activities can be supported by cryptocurrency-valued transactions. 
  Furthermore, cryptocurrencies can be used in various other scenarios in the metaverse such as advertising, e-commerce, event organizing, social networks, etc.

\subsection{Digital Wallets for Metaverse}\label{sec:DigitalWallets}


Since users are the foundation of the metaverse ecosystem, a natural question is how to enable users to interact with the economic systems in metaverse. The related activities include the management of their digital assets and the participation in DeFi activities in metaverse. In this part, we discuss two related questions, i.e.,
\begin{itemize}
    \item What are digital wallets in the real world?
    \item How do users interact with metaverse using their digital wallets?
\end{itemize}

 \subsubsection{Digital Wallets in Real World}
    
    A digital wallet is software that allows users to interact with blockchains. It provides users services such as storing private keys of users' crypto assets, conducting transactions, and invoking smart contracts. In Ethereum, digital wallets \cite{EthereumWallet} are applications that help users connect with their Ethereum accounts. One can read his/her account balance, launch transactions, and connect to dApps using their wallets. MetaMask \cite{metamask} is the most well-known crypto wallet, which can be used to link any crypto's trading platform of the user's choice. Users can deposit their local fiat currency and then convert it for cryptocurrencies through exchanges such as Coinbase and Binance.
    
    According to the way of using private keys, digital wallets can be classified into \textit{software wallets} and \textit{hardware wallets}. Software wallets are further classified into \textit{hot wallets} and \textit{cold wallets} \cite{WalletReview}. A hot wallet is connected to the Internet. Users can use digital assets directly through their web/mobile wallets. A cold wallet is offline to ensure that the private key is never exposed to the Internet. Thus, a cold wallet can effectively prevent hackers from stealing users' private keys to crypto assets.

  \subsubsection{How to Interact with Metaverse using Digital Wallets}
    
    In blockchains, both token transfers between accounts, and interactions between users and smart contracts are driven by transactions. Users' wallets store their public addresses and the corresponding private keys. A user can initiate a transaction signed by his/her private key through the wallet, thus realizing interactions with blockchains.
    
    All digital wallets have the function of balance querying, token receiving, and transferring metaverse tokens. In addition, some famous wallets also provide unique functions for business convenience. For example, both \textit{Trust Wallet} and \textit{MetaMask} provide access to various NFT marketplaces and game assets.
    
    Argent \cite{argent} is a smart-contract wallet controlled by codes instead of a user's private key. Any function of Argent can be realized through smart contracts. For example, Argent provides \textit{Guandians mechanism} to recover a wallet if a user's private key is lost. Moreover, Argent can set transfer limits, approve transfers to known addresses, and lock accounts to prevent crypto assets from getting spent. These functions can lower the threshold for users when they use cryptocurrencies.
    
    Following the development of Ethereum, these mainstream wallets also support transactions on Ethereum's \textit{Layer2} network. Layer2 solutions offer low fees and fast transactions without compromising security. Users can withdraw their crypto assets to a \textit{Layer1} blockchain at any time. Argent has exploited zkSync \cite{zksync} to support its Layer2 wallet.

\section{Economic Activities for Metaverse Users}\label{sec:economicActivities}

    

    
   Decentralized Finance (shortened as DeFi) is a paradigm of financial service technologies based on distributed ledgers built on top of blockchains. DeFi directly provide transaction services to customers without any intermediaries like centralized banks and governments. All DeFi transactions will be recorded in blockchains and become immutable.

 In future metaverse, when users own their digital assets, they would very possibly participate in various economic activities such DeFi protocols, aiming to earn extra profit.
 Given that DeFi boosted the historical second-wave blockchain applications about five years ago, we are curious about the following three questions:
   \begin{itemize}
       \item[i)] What popular DeFi Protocols are there in real world? 
       \item[ii)] What are the differences between the real-world DeFi and that in metaverse?
       \item[iii)] How do users take part in DeFi protocols in metaverse?
   \end{itemize}
   
   In the following, we first review popular DeFi protocols, then discuss how to participate into DeFi activities in mateverse.

 \subsection{Most Popular Real-World DeFi Protocols}
 

\begin{table}[t]
\caption{Popular DeFi Protocols \cite{defillama}}
\centering
\footnotesize
\begin{tabular}{|m{2.1cm}<{\centering}|m{5.7cm}|}%
\hline
\textbf{Name} & \textbf{Recognition} \\
\hline
    DEXes \cite{dexes} & Protocols where users can swap and trade cryptocurrencies.\\
    \hline
    Lending \cite{Lending} & Protocols that allow users to borrow and lend assets.\\
	\hline
	CDP \cite{cdp} & Protocols that mint its own stablecoin using collateralized lending.\\
	\hline
	Liquid Staking \cite{Liquid} & Protocols that allow users to stake assets in exchange for a reward.\\
	\hline
	Bridge \cite{Bridge} & Protocols that bridge tokens from one network to another.\\
	\hline
	Yield \cite{Yield} & Protocols that pay you a reward for your staking/LP on their platform.\\
	\hline
	Services \cite{Services} & Protocols that provide a service to the user.\\
	\hline
	Derivatives \cite{Derivatives} & Protocols for betting with leverage.\\
	\hline
	Yield Aggregator \cite{Yield_Aggregator} & Protocols that aggregated yield from diverse protocols.\\
	\hline
	Algo-Stables \cite{Algo-Stables} & Protocols that provide algorithmic coins to stablecoins.\\
	\hline
	Cross Chain \cite{Cross_Chain} & Protocols that add interoperability between different blockchains.\\
	\hline
	Synthetics \cite{Synthetics} & Protocol that created a tokenized derivative that mimics the value of another asset.\\
	\hline
	Launchpad \cite{Launchpad} & Protocols that launch new projects and coins.\\
	\hline
	Reserve Currency \cite{Reserve_Currency} & OHM forks: Protocols that use a reserve of valuable assets acquired through bonding and staking to issue and back its native token.\\
	\hline
	Insurance \cite{Insurance} & Protocols that are designed to provide monetary protections.\\
	\hline
\end{tabular}
\label{Table:ExistingDeFi}
\end{table}

Table \ref{Table:ExistingDeFi}, presents the 15 most popular categories of DeFi protocols provided by DefiLlama \cite{defillama}. Those DeFi categories are ranked according to their Total Value Locked (TVL), which is one of the key indicators to help us understand the value of a smart-contract DeFi protocol. 
In the following, we take the first 3 categories as examples to show the representative products for each.


 \subsubsection{DEXes}


  Uniswap \cite{Uniswap} is essentially a smart contract-enabled decentralized exchange (DEX), which offers the trading protocol for cryptocurrencies and automatic liquidity. Using the native governance token \textit{UNI}, Uniswap performs as an Automated Market Maker (AMM), which conducts peer-to-peer market making based on an automated trading algorithm, and enables the swapping of multiple ERC-20 tokens on the Ethereum blockchain.


  Curve \cite{curve} is another well-known DEX, mainly using stablecoins and its native governance token called \textit{CRV}. As a liquidity aggregator, Curve pegs users' crypto assets with a specific token such as Bitcoin and stablecoins. Curve is widely exploited for the swap of stablecoins because it guarantees a low slippage rate and a low swap fee of 0.04\%. In addition to providing liquidity, Curve's users can earn transaction fees, and benefit from the interactions with other DeFi protocols like Compound, Synthetix, RenBTC, etc.


  Saddle \cite{Saddle} implements an AMM on the Ethereum blockchain, optimized for trading pegged value crypto assets with minimal slippage. It efficiently supports DeFi trading between stablecoins and its pegged crypto assets such as wETH and wBTC.

 \subsubsection{Lending}


  Compound \cite{Compound} is a lending protocol that supplies crypto assets as collateral in order to borrow the base asset. Compound accounts can earn interest by lending the base asset to the protocol. Meanwhile, its accounts can also acquire governance token called \textit{COMP}, using which accounts can help make critical governance decisions of the protocol network.


  Aave \cite{Aave} is one of the most popular and cutting-edge lending protocols. It realizes a decentralized non-custodial liquidity protocol, in which users can participate as depositors or borrowers. As a depositor, a user can earn passive income by supplying liquidity to the market. While acting as a borrower, a user can get the borrowed assets either by the permanent excess mortgage or low-collateralized flash credit. The native token of Aave is named \textit{AAVE}, which enables users to participate in the governance of its protocol.

 \subsubsection{Collateralized Debt Position (CDP)}


 MakerDAO \cite{markerDAO} is a CDP protocol built on the Ethereum blockchain, dedicated to bringing stability to the cryptocurrency economy. It consists of the stablecoin \textit{DAI}, the maker collateral vault, Oracle, and a voting mechanism. Using its native governance token \textit{MKR}, the token holders can decide the key parameters of the MakerDAO protocol such as the swap rate of stablecoins and the ratio of collateral.\\

 



 

 \subsection{DeFi in Real World v.s. DeFi for Metaverse}

  We now discuss the second question, i.e., what differences are there between the real-world DeFi and that in the metaverse?
  Observing that a lot of popular DeFi protocols have been developed and applied in the real-world crypto market, we are also curious about what DeFi activities and products in the metaverse will look like.

  The future metaverse will definitely require DeFi protocols in the virtual world. Even today, we see that metaverse cryptocurrencies have already been exploited to enable the economic system in some metaverse games such as Roblox \cite{Roblox} and The Second Life \cite{2ndlife}.
  For example, The Second Life enables its players to shop in the built-in marketplace. Players can purchase millions of virtual fashions, home decor, and other items. The Second Life can even allow game players to generate users' own digital creations and monetize them to earn profits in a virtual economy powered by Tilia \cite{Tilia}. Here, Tilia is an all-in-one payment platform dedicated to the metaverse economy. With Tilia, game players, NFT providers, and the publishers of the virtual world can launch their payments, initiate virtual crypto tokens, and earn real assets in both the real world and the metaverse.

 Metaverse enables people to shop, game, buy and trade currencies and other objects. In the metaverse, cryptocurrencies serve as money in a virtual digital world. There exist several terms like ``metaverse coins", ``metaverse tokens" and ``metaverse crypto". Every metaverse project has to handle transactions within a particular digital environment through its native tokens.

  Recently, it is reported that the financial sector of Meta \cite{FaceBookMetaverse}, i.e., Meta Financial Technologies, is exploring a new crypto token, internally named \textit{MarkCoin}, to develop their economic system in Meta's virtual business.
 Numerous other metaverse projects have already been launched, and their individual tokens are available in the crypto market such as MANA \cite{Decentraland}, SAND \cite{Sandbox}, and HIGH \cite{Highstreet}.
 %
 MANA is the native token of Decentraland \cite{Decentraland}. In this metaverse-style game, users can buy and sell virtual land, estates, avatar wearable gadgets, and even names in the Decentraland marketplace. All the transactions while stocking these digital goods are supported by the Ethereum blockchain.
%
 The Sandbox \cite{Sandbox} was founded in 2012 by Pixowl as a mobile gaming platform. In 2021, it was upgraded to a play-to-earn blockchain version and becomes one of the fastest-growing crypto-driven metaverse games. In this new-version game, players can build their virtual worlds using the NFTs backed by Ethereum. For example, using its native token SAND, people can buy a piece of land from The Sandbox metaverse to host their virtual events.
%
 HIGH is another metaverse project named ``Highstreet" \cite{Highstreet}, which is an interesting project with VR-supported metaverse applications. A user can shop for things inside its virtual universe using the currency HIGH.
 
 Via reviewing several real metaverse projects and their corresponding token systems, we see that the DeFi activities in the metaverse keep growing. 
 In the next subsection, we investigate the third question, i.e., how do users take part in DeFi activities in the metaverse?

  We answer this question following two threads, i.e., how to participate into DeFi activities as customers, and how to engage in DeFi in metaverse as developers.
 %
 %
 The first thread is straightforward. As customers of DeFi protocols in the metaverse, the target is to earn \textit{profit}. Thus, the users who own metaverse crypto assets can take part in various DeFi lending protocols aforementioned. However, please note that we are not encouraging anyone to invest any DeFi product. Any economic activity has the risk of failure. Please do your own research before taking actions.

 \subsection{How to Enable DeFi in Metaverse as Developers?}


We take Polygon \cite{Polygon} as an example to present our thoughts.


\subsubsection{What is Polygon}

 Polygon \cite{Polygon} is claimed as a decentralized platform that is committed to bringing the world to Ethereum. Using it, developers can create low-fee dApps without compromising the security of the blockchain. Polygon aims to become a solution aggregator in 
 the Layer2 blockchain of Ethereum and offer a modular, universal, and flexible framework to scale out the service level of Ethereum.

\subsubsection{History of Polygon}

Ploygon has experienced several stages.

\begin{itemize}
    \item \textit{Early stage (2017-2019).} In 2017, Matic Network was born, which is the previous version of Polygon. At the beginning, Matic was positioned as a single Layer2 solution, which adopted the Plasma framework \cite{plasma} as the solution.


 \item \textit{Development stage (2017-2019).} As the project evolved, Matic Network expanded its scope from a single Layer2 solution to a ``network of networks", eventually changing its name to ``Polygon" in February 2021.
   
 \item \textit{Current stage (2021-present).} Currently, Polygon has only one mature product named as \textit{Polygon PoS} \cite{Polygon_Pos}. The Polygon team plans to focus more on \textit{rollup} \cite{optimisticRollups} solutions in the future.

\end{itemize}



 \subsubsection{Polygon PoS}
 
 Polygon PoS \cite{Polygon_Pos} is an extension solution of Ethereum that achieves fast transaction speed and low-cost fees by leveraging sidechains for transaction processing. It also uses Plasma \cite{plasma} as a secure bridging framework, and PoS validators to guarantee the security of the on-chain assets.


\subsubsection{What We can Learn from Polygon}

In the future metaverse, there will be a universal standard that defines the rules of executing DeFi protocols.
The question is that if moving the existing popular DeFi protocols to the metaverse scenarios, people are wondering can they work in the context of metaverse.
We can imagine that metaverse needs a high-throughput and low-cost transaction hub like Polygon. Let us call this hub the \textit{universal hub}, which aims to enable metaverse users to perform frequent interactions with the bottom-layer blockchains of multiple metaverse platforms.

 Given that users' frequent cross-platform interactions of metaverse demand low-latency and high-throughput transactions across multiple metaverse platforms, the next question is what exactly Polygon inspires us to conduct DeFi in the metaverse.
 Firstly, Polygon has addressed the heterogeneity of the diverse blockchains in its bottom layer. 
 Secondly, Polygon enables excellent ability of cross-platform transactions.
 In addition, the Polygon-like universal hub should support developers to implement customized solutions in Layer2 of the bottom blockchain, while considering multiple metrics.

 %
 %

\section{Cross-chain Ecosystem for Metaverse}\label{sec:Cross-chain-Ecosystem}

    


In this section, we first analyze why metaverse needs cross-chain solutions. Based on the existing cross-chain solutions and their pros and cons, combined with the characteristics of the metaverse economic system, we present our cross-chain solutions that are suitable for the metaverse. Through the observation of existing metaverse projects, We found that the behaviors and transaction rules across multiple metaverse platforms cannot be found yet. Therefore, following the cross-chain logic, we finally present a deductive analysis of interoperability across multiple metaverses.

 \subsection{Cross-chain Technologies}
 
 Since the debut of Bitcoin, various blockchains and their applications are been proposed. According to the survey of coinmarketcap \cite{coinmarketcap}, 9,154 cryptocurrencies have been used in the real world, where more than 8,000 independent blockchain systems are involved. Therefore, it is envisioned that \cite{multichain,multichainapproach} the ecosystem will develop a multi-chain future where various blockchain protocols coexist. In the multi-chain ecosystem, users can experience the services of various blockchain systems by joining a blockchain system. Decentralized applications designed by developers are not limited to a single closed system. Instead, dApps intend to create enormous value in a more broad multiple-blockchain ecosystem. A newly deployed blockchain naturally chooses to cooperate with existing mainstream other blockchains and gain their support. The key to supporting multi-chain ecological integration is \textit{cross-chain technologies}, which enable two or more independent blockchains to interoperate with on-chain objects (assets or other associated data).

  Most blockchain ecosystems do not support cross-chain technologies. Gavin Wood, the co-founder of Ethereum, launched a heterogeneous multi-chain blockchain network named \textit{Polkadot} to achieve interoperability across blockchains \cite{Polkadot}.
    Particularly, the Polkadot network consists of multiple \textit{parachains} and a \textit{Relay Chain}. Parachains are used to process transactions in parallel. The Relay Chain is the main chain of the system, providing security guarantees for the Polkadot. Polkadot defines a set of message format standards to provide interoperability for parachains. In order to interact with external blockchains such as Bitcoin and Ethereum, Polkadot uses various \textit{blockchain bridges} to allow external blockchains to share arbitrary data such as crypto assets and NFT.

 \subsection{Metaverse Needs Cross-chain Solutions}
 
 The metaverse is a virtual space that maps from and is independent of the real world. It is not a single closed universe, but a constantly expanding digital universe composed of boundless virtual worlds and digital content. Blockchain technology can provide underlying infrastructure support for the metaverse. That is, the identity and economic system in the metaverse operate based on the blockchain infrastructure, which is independent of a specific operator so that the data is owned by the user and not monopolized by an operator. In particular, blockchain technologies guarantee the uniqueness, privacy, and security mechanisms of avatars in the metaverse. Furthermore, a decentralized economic system in the metaverse requires blockchains to ensure security and trust. In the metaverse, users may conduct a large number of transactions at any time, and those transactions must be verified in a decentralized way. 
In addition, digital assets must be capitalized with blockchains to identify their authority. The trading and circulation of those digital assets rely on the underlying blockchain technology.

The metaverse is essentially a digitized version of the real world. What it builds is a space where the real and virtual worlds are highly integrated and interactive. Meanwhile, the metaverse is also an open, fair, and distributed world. Every individual or organization can build a virtual space even another sub-metaverse in the metaverse. Moreover, the development of the metaverse is not decided by a single company or enterprise, but via multi-party collaboration. Therefore, cross-platform communications between or within the metaverse brings new challenges to user identity uniqueness, asset's transfer, and digital-content's circulation. All those mentioned facts implicate that there is an urgent need to design cross-chain solutions for the metaverse.

 \begin{figure*}[t]
        \centering     \includegraphics[width=0.8\linewidth]{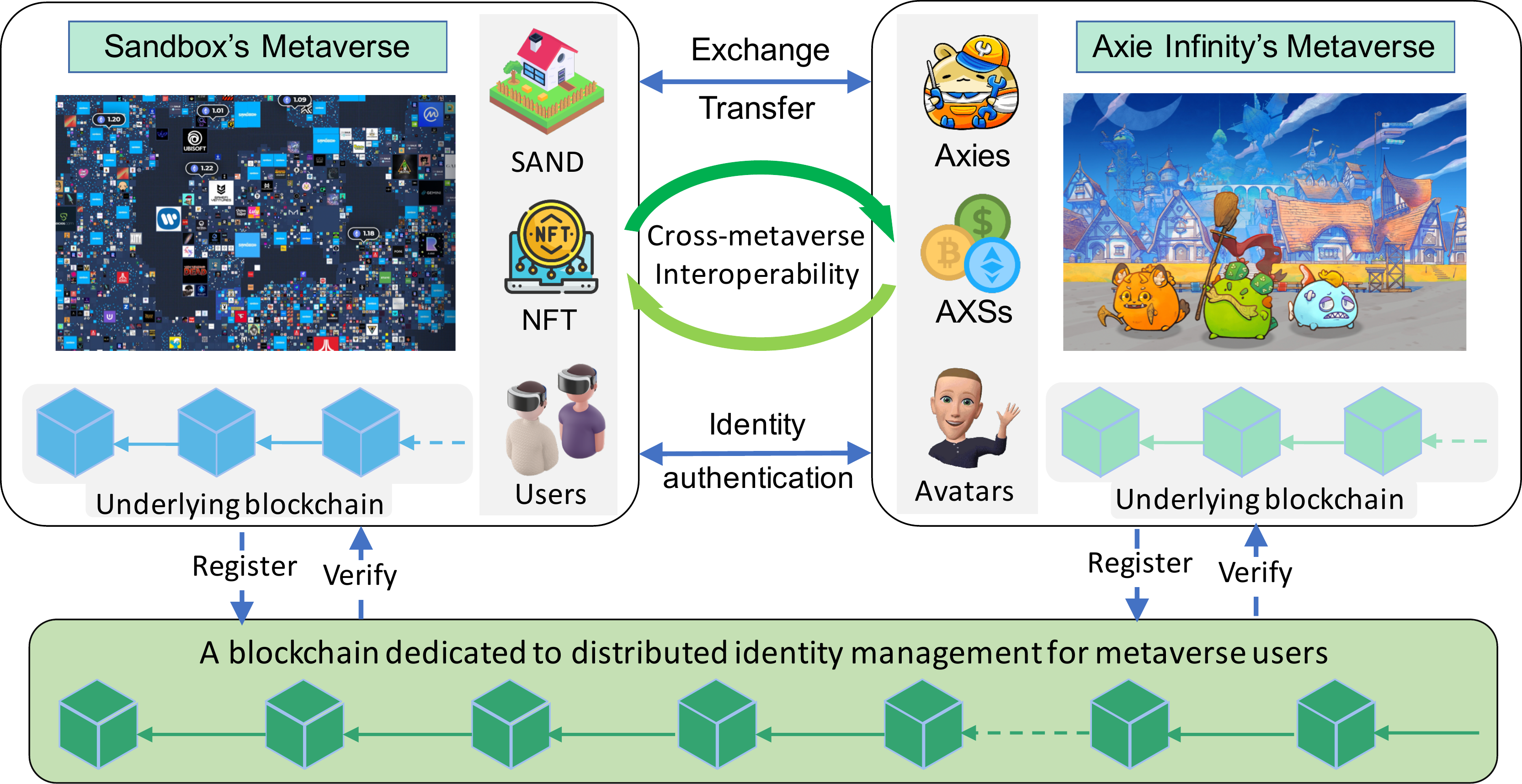}
        \caption{Illustration of cross-metaverse interoperability, in which we take Sandbox's and Axie Infinity's metaverses as examples.}
        \label{fig:cross-chain}
 \end{figure*}

\subsection{Cross-chain Protocols for Metaverse}

Since blockchains are independent of each other, a cross-chain communication protocol that enables blockchains to communicate with each another is required. A typical cross-chain protocol is divided into four phases \cite{zamyatinAZKMKK21}: {\itshape pre-commit} phase, {\itshape verify} phase, {\itshape commit} phase and {\itshape abort} phase. The {\itshape 
pre-commit} and {{\itshape commit} phases are equivalent to locking and unlocking the states of the two blockchains, respectively. The {\itshape verify} phase is a key component of the cross-chain protocol, because it allows one blockchain to perceive the state transition of the other. For example, suppose that we have two blockchains, i.e., chains \textit{X} and \textit{Y}. In order to unlock assets on chain \textit{Y}, consensus nodes of chain \textit{Y} need to verify that the backed assets on chain \textit{X} have been locked. The {\itshape abort} phase means that the {\itshape pre-commit} state reverts to the previous state if the {\itshape verify} or {\itshape commit} phase fails. 

According to the current design principles, cross-chain protocols can be classified into three categories: i) trusted cross-chain protocols}, ii) \textit{trustless cross-chain protocols}, and iii) \textit{hybrid protocols}, as described in Table \ref{Table:cross-chain}.

\begin{table}[t]
\caption{Cross-chain Protocols and Their Representative Examples}
\centering
\footnotesize
\begin{tabular}{|m{2.2cm}<{\centering}|m{5.7cm}|}%
\hline
\textbf{Cross-chain Protocols} & \textbf{Representative Examples} \\
\hline 
    Trusted Cross-chain Protocols & Notary \cite{coinbase}, Tokrex \cite{Tokrex}, Corda \cite{corda}, BTCB \cite{BTCB}, HBTC \cite{HBTC}, tBTC \cite{tBTC}, ren \cite{ren}, DeCus \cite{DeCus}, Hop Exchange \cite{HopExchange}, Hyphen \cite{Hyphen}, Degate Bridge \cite{DegateBridge}.\\
    \hline
    Trustless Cross-chain Protocols & Hash Time-lock \cite{herlihy2018atomic}, WBTC \cite{WBTC}, cBridge \cite{cBridge}, Lightning Network \cite{LightningNetwork}, SEPoW \cite{li2021sepow}, Zcash XCAT \cite{ZcashXCAT}, Interledger \cite{Interledger}, BTCRelay \cite{btcrelay}.\\
     \hline
     Hybrid Protocols & Xclaim \cite{zamyatin2019xclaim}, zkBridge \cite{xie2022zkbridge}, Plasma \cite{plasma}, Polkadot \cite{Polkadot}, Cosmos \cite{cosmos}, Decentralized gateway \cite{ghosh2021leveraging}.\\
     \hline
\end{tabular}
\label{Table:cross-chain}
\end{table}

The representative example of \textit{trusted cross-chain protocol} is Notary \cite{coinbase}. In the {\itshape pre-commit} phase of protocols, a user locks the assets of chain \textit{X} to the notary's committee address, aiming to obtain the corresponding unlocked assets on chain \textit{Y}. In the {\itshape verify} phase, the notary waits for the assets on chain \textit{X} that are to be confirmed. If these assets are confirmed, the protocol enters the {\itshape commit} phase. The notary unlocks the corresponding assets on chain \textit{Y} to the user's address on chain \textit{Y}, then completes the cross-chain transfer. 
We found that in the procedure described above, users' assets are managed and controlled by a notary. In order to become a notary, existing cross-chain protocols have different methods, such as depositing assets, evaluating applicants' reputation value, and selecting consensus committees. 
We then review the pros and cons of Notary protocol.
Firstly, the Notary protocol's pros include: i) efficiency, which can quickly realize the cross-chain transfer of assets, and ii) practicality, which can be better compatible with existing blockchain systems. 
On the other hand, Notary's cons include insecurity and high fees. The cross-chain transfer of assets relies on an external trusted committee. If the committee is malicious, cross-chain assets are vulnerable. The assets deposited by the committee are with high diversity. To balance the committee members' income, a higher cross-chain service fee is charged.

The representative example of \textit{trustless cross-chain protocol} is hash time-lock \cite{herlihy2018atomic}. In this protocol, the {\itshape pre-commit} and {\itshape commit} phases are carried out simultaneously, aiming to lock assets into the counterparty's address. In the {\itshape verify} phase, a user performs the unlocking operation on chain \textit{Y} by revealing the unlocked secret. On the other hand, the counterparty who has obtained the revealed secret performs the unlocking operation on chain \textit{X}, and  completes the exchange of assets. 
In addition, sidechain protocols \cite{btcrelay} can also realize the cross-chain transfer of assets without a trusted third party. The advantage of these protocols is high security and no need an intermediary. However, their disadvantage is that they rely on synchronization assumptions, which entail that cross-chain participants must be online at the same time.

The representative example of \textit{hybrid protocol} is {\itshape Xclaim} \cite{zamyatin2019xclaim}. In this protocol, both trusted and trustless models are exploited. To transfer assets from chain \textit{X} to 
chain \textit{Y}, chain \textit{Y} can verify the validity of transactions that took place on chain \textit{X} without a trusted third party. This is realized by relaying consensus information, such as block headers, on chain \textit{X} to the relay contract resided on chain \textit{Y}. In the opposite direction, the transfer of the assets is completed by a trusted notary.

\subsection{Promising Cross-chain Solutions for Metaverse}

The metaverse is an open and comprehensive virtual world. Every individual and organization can create their own metaverse platforms. They can adopt different technical solutions and design different customized application scenarios. Based on these characteristics, we believe that the following cross-chain solutions are promising for the metaverse.

\subsubsection{Supporting the interoperability of heterogeneous models}

Different metaverses may be built on top of underlying heterogeneous blockchains with different functionalities and crypto algorithms. If the cross-chain protocol does not properly adapt to these heterogeneous blockchain models, it might lead to failures of cross-chain transfer. For example, there exists a cross-chain protocol \cite{btcrelay} in which its implementation depends on smart contracts. However, this protocol is useless for blockchains without equipped with smart contracts. In addition, the hash time-lock protocol requires two interlinked blockchains to use the same hash function. That is, chain \textit{X} uses a function $H(\cdot)$ such that $h=H(x)$. On the other hand, chain \textit{Y} uses a hash function $H'(\cdot)$ such that $h \neq H'(x)$. Therefore, a design challenge in cross-chain protocols is how to achieve interoperability between heterogeneous blockchains.

\subsubsection{Supporting privacy preserving}

Privacy is a pivotal property of metaverse ecosystems, and thus also applies to cross-chain protocols. Ideally, the assets of cross-chain transfer and the identity of the transfers can be hidden from the public. Unfortunately, most cross-chain protocols lack the module for privacy preservation. For example, in the hash time-lock protocol, two transactions that execute asset commits on two blockchains are publicly recorded in the blockchain. By observing these two transactions, all blockchain users know who made these cross-chain transfers and how many tokens were transferred. Therefore, the metaverse should provide a cross-chain protocol that supports identity anonymity and asset-blinding mechanisms.

\subsubsection{Supporting multiple-object interoperability}
   
A cross-chain protocol dedicated to the metaverse should support the interoperability of any information across multiple metaverses, including at least digital assets, instructions of smart contracts, and user identities. Among those economic activities across metaverse platforms, a typical example is the \textit{asset transfer} through underlying blockchains. In order to improve cross-chain performance, it is necessary to realize automatic cross-chain asset transfer through smart contracts. Thus, the instructions of smart contracts across blockchains need to be supported by such cross-chain protocol. In addition, to facilitate users to travel across metaverses, the verification of users' identities across blockchains should be also implemented.

\subsection{How will Cross-Chain Ecosystem Evolve in Metaverse}

In the real world, people can engage in a wide variety of activities. For example, during work hours, people work in their companies for payments. At night, people spend in entertainment venues to enjoy their life with the money earned from their work. Similarly, users of the metaverse also need to travel freely through applications across different metaverse platforms. As shown in Fig. \ref{fig:cross-chain}, as a metaverse user, Alice plays games in Sandbox \cite{Sandbox} and earns crypto assets called \textit{SAND}. Subsequently, Alice wants to participate in activities in Axie Infinity \cite{axieinfinity}. To this end, Alice needs to purchase \textit{Axies} using her \textit{SAND}. From this real-world example, we observe that in order to accomplish Alice’s cross-metaverse activities, two events are involved, i.e., i) the identity verification of cross-metaverse, and ii) the asset transfer across metaverse platforms. From a technical point of view, these two events sparked a cross-metaverse transaction. Since both Sandbox and Axie Infinity are built on top of blockchain technologies, the cross-chain protocol can handle the aforementioned cross-metaverse transactions, Thus, such cross-chain protocol can also support Alice's cross-metaverse activities.
It can be foreseen that the cross-chain protocols will receive growing attention in the context of cross-metaverse applications.

\section{Challenges and Open Issues}\label{sec:openIssues}

Although we have reviewed a lot of related studies, the economic systems dedicated to metaverse are still in its initial stage from the perspective of either industry or academia. In this section, we summarize challenges and open issues that still need to be addressed in the upcoming few years.


\subsection{How to Interpret Authorship and Inventorship of AIGC}


 In the context of accelerated integration of the digital and physical worlds, AIGC is leading a profound change, reshaping the production and consumption modes of digital content. It will greatly enrich people’s digital life, and also is an indispensable force for the future development of digital civilization.
 However, an open issue is how to interpret the \textit{authorship} of creations, and the \textit{inventorship} of inventions generated by AI algorithms \cite{Ballardini2019AIgeneratedCA}. We look forward to seeing new discussions and solutions proposed very soon.


\subsection{Risk of Avatars in Metaverse}
    
 In the human-centric metaverse, the virtual world is composed of many avatars controlled by users. Avatars sense the physical world's state through interactive devices \cite{heller2016avatars}. Users can enter and live in the virtual habitat through the digital avatars, which interact with other virtual entities \cite{genay2021being} to communicate, collaborate, socialize, and work in the virtual world. The interactive activities of virtual and reality require large-scale distributed computing and communication infrastructure, and rely on emerging multimedia technologies such as AR, VR, MR, and tactile internet (TI). However, there are still some inherent risks following the current mode of digital avatars in metaverse. In this regard, we outlook some possible risks related to avatars as follows.


 \subsubsection{Fraud Risks} 
 
 In an open metaverse ecosystem, avatars support the collaborative production-based economy. For example, Roblox \cite{Roblox} allows any user to create games and avatars. However, in this multi-entity creative economy, each participant's credibility cannot be guaranteed \cite{liao2021digital}. Thus, it is a challenge to eliminate malicious frauds when creating and publishing digital entities.

 \subsubsection{Metaverse Crimes} 
 
 The safety of avatars is being threatened by crimes in the metaverse. Such crimes seriously affect the stability of the virtual habitat. Yang {et al.} \cite{yang2022secure} propose a biometric identity authentication framework based on the chameleon signature to identify and track malicious players. In order to guarantee the consistency of the user and the associated avatar, the authors construct an identity model based on biometrics to achieve the verifiability of players' physical identity.

 \subsubsection{Financial Risks}
 
 In metaverse's trading market, the trading of virtual digital products  among multiple entities has inherent financial risks (e.g., refusal to pay). For example, because of a reentrancy flaw in smart contract codes \cite{wang2022survey}, the metaverse project Paraluni based on Binance Smart Chain (BSC) lost over \$1.7 million in 2022.

\subsection{Outlook of Monetary Systems in Metaverse}


 In the future, it will definitely exist multiple metaverses in our daily life. Although a metaverse is possibly built by a giant commercial group, the economic system in the metaverse must be operated based on decentralized token systems. This is because transactions in metaverse are conducted using cryptocurrencies without relying on any third-party intermediary. Due to safety and security reasons, transactions that occurred in the metaverse must be recorded in a blockchain. It can be foreseen that each metaverse very possibly will launch at least a native cryptocurrency, which becomes the virtual economy's legal cash. This paradigm also indicates that dedicated circulating protocols are needed to enable economic activities across metaverses.

  Several representative real-world examples are reviewed here as references to design such metaverse's circulating protocols.
  Tian \textit{et al.} \cite{tian2021enabling} propose a distributed cryptocurrency exchange protocol, in which two types of consensus mechanisms (i.e., Proof of Work and Proof of Deposit) are used to select trustworthy users for constructing a validation committee.
  As an existing successful solution, Uniswap \cite{angeris2019analysis} as a decentralized exchange built on Ethereum automatically provides services for the liquidity of cryptocurrencies.

 Referring to those existing exchange protocols, we could imagine that virtual exchanges will appear in the future metaverse. People will directly trade digital assets including both fungible and non-fungible tokens, and other digital assets via such new DEXes in the virtual world.

 On the other hand, the risks related to virtual DEXes cannot be ignored. Even worse, virtual DEXes are possibly more vulnerable to hackers' attacks in the virtual world than those in physical world. In addition, it might be more difficult to detect some economic scams in the metaverse. For example, Phishing schemes \cite{Wu2020who} and Ponzi Schemes \cite{Chen2018Detecting} can be designed following completely new architectures in the metaverse. It causes difficulty for governance when those new types of scams are widely spread in the virtual world.
 Therefore, how to tackle those new threats toward the future healthier DEXes in the metaverse becomes a promising research direction.

 \subsection{How Cross-Chain Technology Supports Wallet Apps}

    
    The cross-chain wallet is an upgraded version of the conventional digital wallet. The cross-chain wallet is built for interoperable multi-chain-based economic applications. When powered with interoperability, a digital wallet becomes a cross-chain wallet that can connect to multiple blockchain ecosystems \cite{Polkadot}. Users can directly interact with a wide range of cross-chain web3 applications with their unique wallet addresses, eliminating the hassle of maintaining multiple secret keys. 
    
    Cross-chain wallets comprise some unique capabilities that conventional wallets cannot provide. For example, users can transfer their crypto assets and NFTs on different on-chain marketplaces. If someone's digital assets are distributed in different chains, he/she can know the total value of his/her assets through the cross-chain wallet.
    However, cross-chain wallets are still in the very initial development stage. New products and solutions are open to implementation in the upcoming years. Numerous technical challenges are needed to tackle, including privacy preservation of wallet users, security of users' assets, risks of hacking attacks, and of course the adaption to regulation and law issues in different areas.

 \subsection{Cross-Metaverse Interoperability}

  Multiple metaverse platforms are emerging with their unique blockchains, forming their own ecosystems. Different versions of metaverses may focus on either social networking, games with  excellent graphical quality, or working scenarios with technical supports \cite{CrossChainMetaverse}.
  However, those metaverse platforms are mutually independent, and there is no information or value transfer among them. For example, if Meta's and Nvidia's metaverse users intend to interact with each other, the authentication of users' identity is a huge challenge.
  Therefore, it is necessary for metaverse service providers to offer sufficient back-end and interoperability support, aiming to bridge multiple metaverse platforms. As inspired by the solutions to bridge numerous blockchains, cross-chain technology is the key to achieving such cross-metaverse interoperability. We can foresee that the metaverse will develop a multi-chain future and will not be monopolized by Ethereum. Many coexisted metaverses will be supported by numerous interoperable blockchains \cite{CrossChainMetaverse}. 

  At the current stage, almost the services provided by the blockchain-based metaverse are about the circulation of cryptocurrencies and NFT. In the upcoming stage, multiple-platform metaverse technologies will provide more immersive services for metaverse users, e.g., multiple-entity fully perceptive interactions, multiple-mode coordination, and super cities fusing both virtuality and reality  in the metaverse. Those advanced services require stronger interoperability of smart contracts and applications across multiple metaverse ecosystems \cite{CrossChainMetaverse}.

\subsection{Economic Model of DAO in Metaverse}

  In the era of Web3.0, the ubiquitous metaverse can provide a decentralized immersive virtual habitat where users are able to construct a decentralized autonomous ecosystem to mitigate the critical issue of monopolists and dictators in the metaverse. In economic systems of the metaverse, users can create digital content based on the blockchain and participate in a Decentralized Autonomous Organization (DAO), which could be widely used to organize massive users to create digital content collaboratively. In this regard, the economic value of digital assets needs to be shared among all stakeholders. This new paradigm, i.e., DAO, can drive the innovation of the metaverse ecosystem.

  DAO has been believed as a promising organization for web3 participants \cite{Liu2021DAO}. DAO has received real application scenarios from some metaverse games. For example, MANA is a native token platform of Decentraland \cite{Decentraland}, enabling users to buy digital assets, like virtual land, avatars, and other game wearables. In this platform, owners of tokens can be encouraged to form a DAO to get the right to vote for the improvements of the platform.
 Although DAO provides the original spirit of a decentralized world, there is still a very long way to go before DAO can be widely adopted by the metaverse. This is because the economic model of DAO is facing the following several technical challenges.
 \begin{itemize}
     \item When metaverse users form a DAO, they have to design an economic model and issue cryptocurrencies, which can be devoted to the governance of communities. Designing a healthy economic model for a DAO is not easy, because the economic model decides the incentive mechanism of the organization.
     
     \item The most important feature of the economic model of a DAO is to provide high liquidity. However, the token systems of real-world DAOs are difficult to guarantee such high liquidity.
     
     \item Another technical challenge when designing the economic model of a DAO is how to ensure the profit can be delivered to core contributors of the community.   
 \end{itemize}

 Anyway, the economic model of DAOs plays a crucial role for metaverse users in the future. This topic will attract growing research attention from both academia and industry.


\section{Conclusion}\label{sec:conclusion}

The economic system is the foundation of the metaverse. Through this article, we mainly introduce the preliminaries, basics of economic systems, the fundamental economic activities, and challenges and open issues of the metaverse. We wish this article can inspire researchers, engineers, and educators to explore more paradigms, products, and dApps for a better future metaverse.

\bibliographystyle{IEEEtran}
\bibliography{Reference}

\begin{thebibliography}{100}
\providecommand{\url}[1]{#1}
\csname url@samestyle\endcsname
\providecommand{\newblock}{\relax}
\providecommand{\bibinfo}[2]{#2}
\providecommand{\BIBentrySTDinterwordspacing}{\spaceskip=0pt\relax}
\providecommand{\BIBentryALTinterwordstretchfactor}{4}
\providecommand{\BIBentryALTinterwordspacing}{\spaceskip=\fontdimen2\font plus
\BIBentryALTinterwordstretchfactor\fontdimen3\font minus
  \fontdimen4\font\relax}
\providecommand{\BIBforeignlanguage}[2]{{%
\expandafter\ifx\csname l@#1\endcsname\relax
\typeout{** WARNING: IEEEtran.bst: No hyphenation pattern has been}%
\typeout{** loaded for the language `#1'. Using the pattern for}%
\typeout{** the default language instead.}%
\else
\language=\csname l@#1\endcsname
\fi
#2}}
\providecommand{\BIBdecl}{\relax}
\BIBdecl

\bibitem{yang2022fusing}
Q.~Yang, Y.~Zhao, H.~Huang, Z.~Xiong, J.~Kang, and Z.~Zheng, ``Fusing
  blockchain and ai with metaverse: A survey,'' \emph{IEEE Open Journal of the
  Computer Society}, vol.~3, pp. 122--136, 2022.

\bibitem{tlili2022metaverse}
A.~Tlili, R.~Huang, B.~Shehata, D.~Liu, J.~Zhao, A.~H.~S. Metwally, H.~Wang,
  M.~Denden, A.~Bozkurt, L.-H. Lee \emph{et~al.}, ``Is metaverse in education a
  blessing or a curse: a combined content and bibliometric analysis,''
  \emph{Smart Learning Environments}, vol.~9, no.~1, pp. 1--31, 2022.

\bibitem{nakamoto2008bitcoin}
\BIBentryALTinterwordspacing
S.~Nakamoto, ``Bitcoin: A peer-to-peer electronic cash system,'' Tech. Rep.,
  2008. [Online]. Available: \url{https://bitcoin.org/bitcoin.pdf}
\BIBentrySTDinterwordspacing

\bibitem{chen2022digital}
C.~Chen, L.~Zhang, Y.~Li, T.~Liao, S.~Zhao, Z.~Zheng, H.~Huang, and J.~Wu,
  ``When digital economy meets web 3.0: Applications and challenges,''
  \emph{IEEE Open Journal of the Computer Society}, 2022.

\bibitem{MetaverseBlockchain}
\BIBentryALTinterwordspacing
N.~Garg, ``{What is Metaverse in Blockchain? And why does it Matter?}'' 2022.
  [Online]. Available:
  \url{https://www.brsoftech.com/blog/metaverse-in-blockchain/}
\BIBentrySTDinterwordspacing

\bibitem{CompaniesMetaverse}
\BIBentryALTinterwordspacing
G.~Weston, ``{Top Tech Firms Investing In Web 3.0},'' 2022. [Online].
  Available:
  \url{https://101blockchains.com/top-tech-firms-investing-in-web-3-0/}
\BIBentrySTDinterwordspacing

\bibitem{Roblox}
\BIBentryALTinterwordspacing
``{Roblox},'' 2022. [Online]. Available: \url{https://corp.roblox.com/}
\BIBentrySTDinterwordspacing

\bibitem{GameMetaverse}
\BIBentryALTinterwordspacing
``{Facebook wants to lean into the metaverse. Here's what it is and how it will
  work},'' 2021. [Online]. Available:
  \url{https://www.npr.org/2021/10/28/1050280500/what-metaverse-is-and-how-it-will-work}
\BIBentrySTDinterwordspacing

\bibitem{van2006strategic}
A.~J. van Niekerk, ``The strategic management of media assets: A methodological
  approach,'' in \emph{Allied Academies, New Orleans Congress}, 2006.

\bibitem{toygar2013new}
A.~Toygar, C.~Rohm~Jr, and J.~Zhu, ``A new asset type: digital assets,''
  \emph{Journal of International Technology and Information Management},
  vol.~22, no.~4, p.~7, 2013.

\bibitem{lee2021creators}
L.-H. Lee, Z.~Lin, R.~Hu, Z.~Gong, A.~Kumar, T.~Li, S.~Li, and P.~Hui, ``When
  creators meet the metaverse: A survey on computational arts,'' \emph{arXiv
  preprint arXiv:2111.13486}, 2021.

\bibitem{duan2021metaverse}
H.~Duan, J.~Li, S.~Fan, Z.~Lin, X.~Wu, and W.~Cai, ``Metaverse for social good:
  A university campus prototype,'' in \emph{Proceedings of the 29th ACM
  International Conference on Multimedia}, 2021, pp. 153--161.

\bibitem{timoshenko2019identifying}
A.~Timoshenko and J.~R. Hauser, ``Identifying customer needs from
  user-generated content,'' \emph{Marketing Science}, vol.~38, no.~1, pp.
  1--20, 2019.

\bibitem{kim2012institutionalization}
J.~Kim, ``The institutionalization of youtube: From user-generated content to
  professionally generated content,'' \emph{Media, culture \& society},
  vol.~34, no.~1, pp. 53--67, 2012.

\bibitem{kobis2021artificial}
N.~K{\"o}bis and L.~D. Mossink, ``Artificial intelligence versus maya angelou:
  Experimental evidence that people cannot differentiate ai-generated from
  human-written poetry,'' \emph{Computers in human behavior}, vol. 114, p.
  106553, 2021.

\bibitem{wang2022survey}
Y.~Wang, Z.~Su, N.~Zhang, R.~Xing, D.~Liu, T.~H. Luan, and X.~Shen, ``A survey
  on metaverse: Fundamentals, security, and privacy,'' \emph{IEEE
  Communications Surveys \& Tutorials}, 2022.

\bibitem{lyytinen2021metahuman}
K.~Lyytinen, J.~V. Nickerson, and J.~L. King, ``Metahuman systems= humans+
  machines that learn,'' \emph{Journal of Information Technology}, vol.~36,
  no.~4, pp. 427--445, 2021.

\bibitem{singer2022make}
U.~Singer, A.~Polyak, T.~Hayes, X.~Yin, J.~An, S.~Zhang, Q.~Hu, H.~Yang,
  O.~Ashual, O.~Gafni \emph{et~al.}, ``Make-a-video: Text-to-video generation
  without text-video data,'' \emph{arXiv preprint arXiv:2209.14792}, 2022.

\bibitem{Ho2022ImagenVH}
J.~Ho, W.~Chan, C.~Saharia, J.~Whang, R.~Gao, A.~A. Gritsenko, D.~P. Kingma,
  B.~Poole, M.~Norouzi, D.~J. Fleet, and T.~Salimans, ``Imagen video: High
  definition video generation with diffusion models,'' \emph{ArXiv}, vol.
  abs/2210.02303, 2022.

\bibitem{Feng2022ERNIEViLG2I}
Z.~Feng, Z.~Zhang, X.~Yu, Y.~Fang, L.~Li, X.~Chen, Y.~Lu, J.~Liu, W.~Yin,
  S.~Feng, Y.~Sun, H.~Tian, H.~Wu, and H.~Wang, ``Ernie-vilg 2.0: Improving
  text-to-image diffusion model with knowledge-enhanced
  mixture-of-denoising-experts,'' \emph{ArXiv}, vol. abs/2210.15257, 2022.

\bibitem{Dong2022DreamArtistTC}
Z.~Dong, P.~Wei, and L.~Lin, ``Dreamartist: Towards controllable one-shot
  text-to-image generation via contrastive prompt-tuning,'' \emph{ArXiv}, vol.
  abs/2211.11337, 2022.

\bibitem{wu2020investigating}
Y.~Wu, Y.~Mou, Z.~Li, and K.~Xu, ``Investigating american and chinese
  subjects’ explicit and implicit perceptions of ai-generated artistic
  work,'' \emph{Computers in Human Behavior}, vol. 104, p. 106186, 2020.

\bibitem{latar2015robot}
N.~L. Latar, ``The robot journalist in the age of social physics: The end of
  human journalism?'' in \emph{The new world of transitioned media}.\hskip 1em
  plus 0.5em minus 0.4em\relax Springer, 2015, pp. 65--80.

\bibitem{xu2022full}
M.~Xu, W.~C. Ng, W.~Y.~B. Lim, J.~Kang, Z.~Xiong, D.~Niyato, Q.~Yang, X.~S.
  Shen, and C.~Miao, ``A full dive into realizing the edge-enabled metaverse:
  Visions, enabling technologies, and challenges,'' \emph{IEEE Communications
  Surveys \& Tutorials}, 2022.

\bibitem{lin2022dai}
M.-B. Lin, B.~Wang, F.~Y. Bocart, C.~Hafner, and W.~K. H{\"a}rdle, ``Dai
  digital art index: a robust price index for heterogeneous digital assets,''
  Tech. Rep., 2022.

\bibitem{niemeyer2015maker}
D.~J. Niemeyer and H.~R. Gerber, ``Maker culture and minecraft: Implications
  for the future of learning,'' \emph{Educational Media International},
  vol.~52, no.~3, pp. 216--226, 2015.

\bibitem{axieinfinity}
\BIBentryALTinterwordspacing
``{Axie infinity},'' 2022. [Online]. Available: \url{https://axieinfinity.com/}
\BIBentrySTDinterwordspacing

\bibitem{wang2021non}
Q.~Wang, R.~Li, Q.~Wang, and S.~Chen, ``Non-fungible token (nft): Overview,
  evaluation, opportunities and challenges,'' \emph{arXiv preprint
  arXiv:2105.07447}, 2021.

\bibitem{lim2022realizing}
W.~Y.~B. Lim, Z.~Xiong, D.~Niyato, X.~Cao, C.~Miao, S.~Sun, and Q.~Yang,
  ``Realizing the metaverse with edge intelligence: A match made in heaven,''
  \emph{arXiv preprint arXiv:2201.01634}, 2022.

\bibitem{chen2022economics}
Y.~Chen and H.~Cheng, ``The economics of the metaverse: A comparison with the
  real economy,'' \emph{Metaverse}, vol.~3, no.~1, p.~19, 2022.

\bibitem{8501910}
H.~R. Hasan and K.~Salah, ``Proof of delivery of digital assets using
  blockchain and smart contracts,'' \emph{IEEE Access}, vol.~6, pp.
  65\,439--65\,448, 2018.

\bibitem{angeris2019analysis}
G.~Angeris, H.-T. Kao, R.~Chiang, C.~Noyes, and T.~Chitra, ``An analysis of
  uniswap markets,'' \emph{arXiv preprint arXiv:1911.03380}, 2019.

\bibitem{daian2020flash}
P.~Daian, S.~Goldfeder, T.~Kell, Y.~Li, X.~Zhao, I.~Bentov, L.~Breidenbach, and
  A.~Juels, ``Flash boys 2.0: Frontrunning in decentralized exchanges, miner
  extractable value, and consensus instability,'' in \emph{2020 IEEE Symposium
  on Security and Privacy (SP)}.\hskip 1em plus 0.5em minus 0.4em\relax IEEE,
  2020, pp. 910--927.

\bibitem{MetaverseDefine}
\BIBentryALTinterwordspacing
smita.verma, ``{A Comprehensive Guide To Building A Metaverse DApp Using
  Unity},'' 2022. [Online]. Available:
  \url{https://www.blockchain-council.org/metaverse/a-comprehensive-guide-to-building-a-metaverse-dapp-using-unity/}
\BIBentrySTDinterwordspacing

\bibitem{zhou2020solutions}
Q.~Zhou, H.~Huang, Z.~Zheng, and J.~Bian, ``Solutions to scalability of
  blockchain: A survey,'' \emph{Ieee Access}, vol.~8, pp. 16\,440--16\,455,
  2020.

\bibitem{FaceBookMetaverse}
\BIBentryALTinterwordspacing
wtflea, ``{Facebook Metaverse: Will it Support Blockchain?}'' 2021. [Online].
  Available:
  \url{http://www.itedge.cn/2021/11/22/facebook-metaverse-will-it-support-blockchain/}
\BIBentrySTDinterwordspacing

\bibitem{solana}
\BIBentryALTinterwordspacing
``{Solana},'' 2022. [Online]. Available: \url{http://solana.io/}
\BIBentrySTDinterwordspacing

\bibitem{Avalanche}
\BIBentryALTinterwordspacing
``{Avalanche},'' 2022. [Online]. Available: \url{http://avax.network}
\BIBentrySTDinterwordspacing

\bibitem{Linera}
\BIBentryALTinterwordspacing
``{Linera},'' 2022. [Online]. Available: \url{https://linera.io/}
\BIBentrySTDinterwordspacing

\bibitem{Sui}
\BIBentryALTinterwordspacing
``{Sui},'' 2022. [Online]. Available: \url{https://sui.io/}
\BIBentrySTDinterwordspacing

\bibitem{Aptos}
\BIBentryALTinterwordspacing
``{Aptos},'' 2022. [Online]. Available: \url{https://aptoslabs.com/}
\BIBentrySTDinterwordspacing

\bibitem{huang2022brokerchain}
H.~Huang, X.~Peng, J.~Zhan, S.~Zhang, Y.~Lin, Z.~Zheng, and S.~Guo,
  ``Brokerchain: A cross-shard blockchain protocol for account/balance-based
  state sharding,'' in \emph{Proc. of IEEE Conference on Computer
  Communications (INFOCOM)}, 2022.

\bibitem{fehr2000fairness}
E.~Fehr and K.~M. Schmidt, ``Fairness, incentives, and contractual choices,''
  \emph{European Economic Review}, vol.~44, no. 4-6, pp. 1057--1068, 2000.

\bibitem{yu2020fairness}
H.~Yu, Z.~Liu, Y.~Liu, T.~Chen, M.~Cong, X.~Weng, D.~Niyato, and Q.~Yang, ``A
  fairness-aware incentive scheme for federated learning,'' in
  \emph{Proceedings of the AAAI/ACM Conference on AI, Ethics, and Society},
  2020, pp. 393--399.

\bibitem{gao2015survey}
H.~Gao, C.~H. Liu, W.~Wang, J.~Zhao, Z.~Song, X.~Su, J.~Crowcroft, and K.~K.
  Leung, ``A survey of incentive mechanisms for participatory sensing,''
  \emph{IEEE Communications Surveys \& Tutorials}, vol.~17, no.~2, pp.
  918--943, 2015.

\bibitem{zhu2016fair}
X.~Zhu, J.~An, M.~Yang, L.~Xiang, Q.~Yang, and X.~Gui, ``A fair incentive
  mechanism for crowdsourcing in crowd sensing,'' \emph{IEEE Internet of Things
  Journal}, vol.~3, no.~6, pp. 1364--1372, 2016.

\bibitem{sinha2017incentive}
A.~Sinha and A.~Anastasopoulos, ``Incentive mechanisms for fairness among
  strategic agents,'' \emph{IEEE Journal on Selected Areas in Communications},
  vol.~35, no.~2, pp. 288--301, 2017.

\bibitem{li2019considering}
D.~Li, L.~Yang, J.~Liu, and H.~Liu, ``Considering decoy effect and fairness
  preference: An incentive mechanism for crowdsensing,'' \emph{IEEE Internet of
  Things Journal}, vol.~6, no.~5, pp. 8835--8852, 2019.

\bibitem{zhu2021impact}
B.~Zhu, W.~Leon, L.~Paul, and P.~Gao, ``Impact of crowdsourcee’s vertical
  fairness concern on the crowdsourcing knowledge sharing behavior and its
  incentive mechanism,'' \emph{Journal of Systems Science and Complexity},
  vol.~34, no.~3, pp. 1102--1120, 2021.

\bibitem{martin2005credibility}
G.~Mart{\'\i}n-Herr{\'a}n and G.~Zaccour, ``Credibility of incentive
  equilibrium strategies in linear-state differential games,'' \emph{Journal of
  Optimization Theory and Applications}, vol. 126, no.~2, pp. 367--389, 2005.

\bibitem{xu2021besifl}
Y.~Xu, Z.~Lu, K.~Gai, Q.~Duan, J.~Lin, J.~Wu, and K.-K.~R. Choo, ``Besifl:
  Blockchain empowered secure and incentive federated learning paradigm in
  iot,'' \emph{IEEE Internet of Things Journal}, 2021.

\bibitem{zhang2015truthful}
X.~Zhang, G.~Xue, R.~Yu, D.~Yang, and J.~Tang, ``Truthful incentive mechanisms
  for crowdsourcing,'' in \emph{2015 IEEE Conference on Computer Communications
  (INFOCOM)}.\hskip 1em plus 0.5em minus 0.4em\relax IEEE, 2015, pp.
  2830--2838.

\bibitem{he2019truthful}
J.~He, D.~Zhang, Y.~Zhou, and Y.~Zhang, ``A truthful online mechanism for
  collaborative computation offloading in mobile edge computing,'' \emph{IEEE
  Transactions on Industrial Informatics}, vol.~16, no.~7, pp. 4832--4841,
  2019.

\bibitem{wang2017towards}
X.~Wang, X.~Chen, and W.~Wu, ``Towards truthful auction mechanisms for task
  assignment in mobile device clouds,'' in \emph{Proc. of IEEE Conference on
  Computer Communications (INFOCOM)}, 2017, pp. 1--9.

\bibitem{liwang2018truthful}
M.~Liwang, S.~Dai, Z.~Gao, Y.~Tang, and H.~Dai, ``A truthful reverse-auction
  mechanism for computation offloading in cloud-enabled vehicular network,''
  \emph{IEEE Internet of Things Journal}, vol.~6, no.~3, pp. 4214--4227, 2018.

\bibitem{ott1965budget}
D.~J. Ott and A.~F. Ott, ``Budget balance and equilibrium income,'' \emph{The
  Journal of Finance}, vol.~20, no.~1, pp. 71--77, 1965.

\bibitem{tang2021incentive}
M.~Tang and V.~W. Wong, ``An incentive mechanism for cross-silo federated
  learning: A public goods perspective,'' in \emph{IEEE INFOCOM 2021-IEEE
  Conference on Computer Communications}.\hskip 1em plus 0.5em minus
  0.4em\relax IEEE, 2021, pp. 1--10.

\bibitem{yang2013truthful}
D.~Yang, X.~Fang, and G.~Xue, ``Truthful incentive mechanisms for k-anonymity
  location privacy,'' in \emph{2013 Proceedings IEEE INFOCOM}.\hskip 1em plus
  0.5em minus 0.4em\relax IEEE, 2013, pp. 2994--3002.

\bibitem{gode1993allocative}
D.~K. Gode and S.~Sunder, ``Allocative efficiency of markets with
  zero-intelligence traders: Market as a partial substitute for individual
  rationality,'' \emph{Journal of political economy}, vol. 101, no.~1, pp.
  119--137, 1993.

\bibitem{myerson1979incentive}
R.~B. Myerson, ``Incentive compatibility and the bargaining problem,''
  \emph{Econometrica: journal of the Econometric Society}, pp. 61--73, 1979.

\bibitem{wang2016fair}
T.~Wang, Y.~Xu, C.~Withanage, L.~Lan, S.~D. Ahipa{\c{s}}ao{\u{g}}lu, and C.~A.
  Courcoubetis, ``A fair and budget-balanced incentive mechanism for energy
  management in buildings,'' \emph{IEEE Transactions on Smart Grid}, vol.~9,
  no.~4, pp. 3143--3153, 2016.

\bibitem{ma2014incentive}
J.~Ma, J.~Deng, L.~Song, and Z.~Han, ``Incentive mechanism for demand side
  management in smart grid using auction,'' \emph{IEEE Transactions on Smart
  Grid}, vol.~5, no.~3, pp. 1379--1388, 2014.

\bibitem{jin2015auction}
A.-L. Jin, W.~Song, P.~Wang, D.~Niyato, and P.~Ju, ``Auction mechanisms toward
  efficient resource sharing for cloudlets in mobile cloud computing,''
  \emph{IEEE Transactions on Services Computing}, vol.~9, no.~6, pp. 895--909,
  2015.

\bibitem{thaler1988anomalies}
R.~H. Thaler, ``Anomalies: The winner's curse,'' \emph{Journal of economic
  perspectives}, vol.~2, no.~1, pp. 191--202, 1988.

\bibitem{loosemore2015inter}
M.~Loosemore and B.~Lim, ``Inter-organizational unfairness in the construction
  industry,'' \emph{Construction management and economics}, vol.~33, no.~4, pp.
  310--326, 2015.

\bibitem{Jiao2018Welfare}
Y.~{Jiao}, P.~{Wang}, D.~{Niyato}, and Z.~{Xiong}, ``Social welfare
  maximization auction in edge computing resource allocation for mobile
  blockchain,'' in \emph{IEEE International Conference on Communications
  (ICC)}, 2018, pp. 1--6.

\bibitem{zhang2021privacy}
M.~Zhang, L.~Yang, S.~He, M.~Li, and J.~Zhang, ``Privacy-preserving data
  aggregation for mobile crowdsensing with externality: An auction approach,''
  \emph{IEEE/ACM Transactions on Networking}, vol.~29, no.~3, pp. 1046--1059,
  2021.

\bibitem{fan2020hybrid}
S.~Fan, H.~Zhang, Y.~Zeng, and W.~Cai, ``Hybrid blockchain-based resource
  trading system for federated learning in edge computing,'' \emph{IEEE
  Internet of Things Journal}, vol.~8, no.~4, pp. 2252--2264, 2020.

\bibitem{xu2021wireless}
M.~Xu, D.~Niyato, J.~Kang, Z.~Xiong, C.~Miao, and D.~I. Kim, ``Wireless
  edge-empowered metaverse: A learning-based incentive mechanism for virtual
  reality,'' \emph{arXiv preprint arXiv:2111.03776}, 2021.

\bibitem{xu2022epvisa}
M.~Xu, D.~Niyato, B.~Wright, H.~Zhang, J.~Kang, Z.~Xiong, S.~Mao, and Z.~Han,
  ``Epvisa: Efficient auction design for real-time physical-virtual
  synchronization in the metaverse,'' \emph{arXiv preprint arXiv:2211.06838},
  2022.

\bibitem{zhang2022truthful}
J.~Zhang, M.~Zong, and W.~Li, ``A truthful mechanism for multibase station
  resource allocation in metaverse digital twin framework,'' \emph{Processes},
  vol.~10, no.~12, p. 2601, 2022.

\bibitem{Luong2018Auction}
N.~C. {Luong}, Z.~{Xiong}, P.~{Wang}, and D.~{Niyato}, ``Optimal auction for
  edge computing resource management in mobile blockchain networks: A deep
  learning approach,'' in \emph{IEEE International Conference on Communications
  (ICC)}, 2018, pp. 1--6.

\bibitem{ng2021double}
J.~S. Ng, W.~Y.~B. Lim, Z.~Xiong, D.~Niyato, C.~Leung, and C.~Miao, ``A double
  auction mechanism for resource allocation in coded vehicular edge
  computing,'' \emph{IEEE Transactions on Vehicular Technology}, vol.~71,
  no.~2, pp. 1832--1845, 2021.

\bibitem{kim2022auction}
S.~Kim, ``Auction, learning and bargaining based control scheme for edge
  assisted metaverse system,'' \emph{Computer Networks}, p. 109462, 2022.

\bibitem{wang2019privacy}
Q.~Wang, J.~Huang, Y.~Chen, X.~Tian, and Q.~Zhang, ``Privacy-preserving and
  truthful double auction for heterogeneous spectrum,'' \emph{IEEE/ACM
  Transactions on Networking}, vol.~27, no.~2, pp. 848--861, 2019.

\bibitem{liew2022economics}
Z.~Q. Liew, Y.~Cheng, W.~Y.~B. Lim, D.~Niyato, C.~Miao, and S.~Sun, ``Economics
  of semantic communication system in wireless powered internet of things,'' in
  \emph{ICASSP 2022-2022 IEEE International Conference on Acoustics, Speech and
  Signal Processing (ICASSP)}.\hskip 1em plus 0.5em minus 0.4em\relax IEEE,
  2022, pp. 8637--8641.

\bibitem{paillier1999public}
P.~Paillier, ``Public-key cryptosystems based on composite degree residuosity
  classes,'' in \emph{International conference on the theory and applications
  of cryptographic techniques}.\hskip 1em plus 0.5em minus 0.4em\relax
  Springer, 1999, pp. 223--238.

\bibitem{jiang2022reliable}
Y.~Jiang, J.~Kang, D.~Niyato, X.~Ge, Z.~Xiong, C.~Miao, and X.~Shen, ``Reliable
  distributed computing for metaverse: A hierarchical game-theoretic
  approach,'' \emph{IEEE Transactions on Vehicular Technology}, 2022.

\bibitem{liu2022incentive}
Y.~Liu, Z.~Fang, M.~H. Cheung, W.~Cai, and J.~Huang, ``An incentive mechanism
  for sustainable blockchain storage,'' \emph{IEEE/ACM Transactions on
  Networking}, 2022.

\bibitem{huang2022joint}
X.~Huang, W.~Zhong, J.~Nie, Q.~Hu, Z.~Xiong, J.~Kang, and T.~Q. Quek, ``Joint
  user association and resource pricing for metaverse: Distributed and
  centralized approaches,'' \emph{arXiv preprint arXiv:2208.06770}, 2022.

\bibitem{sun2021dynamic}
W.~Sun, P.~Wang, N.~Xu, G.~Wang, and Y.~Zhang, ``Dynamic digital twin and
  distributed incentives for resource allocation in aerial-assisted internet of
  vehicles,'' \emph{IEEE Internet of Things Journal}, vol.~9, no.~8, pp.
  5839--5852, 2021.

\bibitem{jiang2021reliable}
Y.~Jiang, J.~Kang, D.~Niyato, X.~Ge, Z.~Xiong, and C.~Miao, ``Reliable coded
  distributed computing for metaverse services: Coalition formation and
  incentive mechanism design,'' \emph{arXiv preprint arXiv:2111.10548}, 2021.

\bibitem{daniel2022ipfs}
E.~Daniel and F.~Tschorsch, ``Ipfs and friends: A qualitative comparison of
  next generation peer-to-peer data networks,'' \emph{IEEE Communications
  Surveys \& Tutorials}, vol.~24, no.~1, pp. 31--52, 2022.

\bibitem{shams2013basics}
F.~Shams and M.~Luise, ``Basics of coalitional games with applications to
  communications and networking,'' \emph{EURASIP Journal on Wireless
  Communications and Networking}, vol. 2013, no.~1, pp. 1--20, 2013.

\bibitem{luo2020incentive}
S.~Luo, X.~Chen, Z.~Zhou, X.~Chen, and W.~Wu, ``Incentive-aware micro computing
  cluster formation for cooperative fog computing,'' \emph{IEEE Transactions on
  Wireless Communications}, vol.~19, no.~4, pp. 2643--2657, 2020.

\bibitem{pu2016d2d}
L.~Pu, X.~Chen, J.~Xu, and X.~Fu, ``D2d fogging: An energy-efficient and
  incentive-aware task offloading framework via network-assisted d2d
  collaboration,'' \emph{IEEE Journal on Selected Areas in Communications},
  vol.~34, no.~12, pp. 3887--3901, 2016.

\bibitem{kang2022blockchain}
J.~Kang, D.~Ye, J.~Nie, J.~Xiao, X.~Deng, S.~Wang, Z.~Xiong, R.~Yu, and
  D.~Niyato, ``Blockchain-based federated learning for industrial metaverses:
  Incentive scheme with optimal aoi,'' in \emph{2022 IEEE International
  Conference on Blockchain (Blockchain)}.\hskip 1em plus 0.5em minus
  0.4em\relax IEEE, 2022, pp. 71--78.

\bibitem{du2022attention}
H.~Du, J.~Liu, D.~Niyato, J.~Kang, Z.~Xiong, J.~Zhang, and D.~I. Kim,
  ``Attention-aware resource allocation and qoe analysis for metaverse xurllc
  services,'' \emph{arXiv preprint arXiv:2208.05438}, 2022.

\bibitem{wang2022infedge}
X.~Wang, Y.~Zhao, C.~Qiu, Z.~Liu, J.~Nie, and V.~C. Leung, ``Infedge: A
  blockchain-based incentive mechanism in hierarchical federated learning for
  end-edge-cloud communications,'' \emph{IEEE Journal on Selected Areas in
  Communications}, 2022.

\bibitem{liu2021blockchain}
H.~Liu, S.~Zhang, P.~Zhang, X.~Zhou, X.~Shao, G.~Pu, and Y.~Zhang, ``Blockchain
  and federated learning for collaborative intrusion detection in vehicular
  edge computing,'' \emph{IEEE Transactions on Vehicular Technology}, vol.~70,
  no.~6, pp. 6073--6084, 2021.

\bibitem{wang2022semantic}
J.~Wang, H.~Du, Z.~Tian, D.~Niyato, J.~Kang \emph{et~al.}, ``Semantic-aware
  sensing information transmission for metaverse: A contest theoretic
  approach,'' \emph{arXiv preprint arXiv:2211.12783}, 2022.

\bibitem{shen2021holistic}
X.~Shen, J.~Gao, W.~Wu, M.~Li, C.~Zhou, and W.~Zhuang, ``Holistic network
  virtualization and pervasive network intelligence for 6g,'' \emph{IEEE
  Communications Surveys \& Tutorials}, vol.~24, no.~1, pp. 1--30, 2021.

\bibitem{han2022dynamic}
Y.~Han, D.~Niyato, C.~Leung, C.~Miao, and D.~I. Kim, ``A dynamic resource
  allocation framework for synchronizing metaverse with iot service and data,''
  in \emph{ICC 2022-IEEE International Conference on Communications}.\hskip 1em
  plus 0.5em minus 0.4em\relax IEEE, 2022, pp. 1196--1201.

\bibitem{lin2021stochastic}
X.~Lin, J.~Wu, J.~Li, W.~Yang, and M.~Guizani, ``Stochastic digital-twin
  service demand with edge response: An incentive-based congestion control
  approach,'' \emph{IEEE Transactions on Mobile Computing}, 2021.

\bibitem{liu2020economics}
Y.~Liu, Z.~Fang, M.~H. Cheung, W.~Cai, and J.~Huang, ``Economics of blockchain
  storage,'' in \emph{ICC 2020-2020 IEEE International Conference on
  Communications (ICC)}.\hskip 1em plus 0.5em minus 0.4em\relax IEEE, 2020, pp.
  1--6.

\bibitem{shen2019equity}
B.~Shen, J.~Guo, and W.~Tan, ``An equity-based incentive mechanism for
  decentralized virtual world content storage,'' in \emph{International
  Conference on e-Business Engineering}.\hskip 1em plus 0.5em minus 0.4em\relax
  Springer, 2019, pp. 19--32.

\bibitem{lin2019making}
X.~Lin, J.~Li, J.~Wu, H.~Liang, and W.~Yang, ``Making knowledge tradable in
  edge-ai enabled iot: A consortium blockchain-based efficient and incentive
  approach,'' \emph{IEEE Transactions on Industrial Informatics}, vol.~15,
  no.~12, pp. 6367--6378, 2019.

\bibitem{hou2021incentive}
W.~Hou, H.~Wen, N.~Zhang, J.~Wu, W.~Lei, and R.~Zhao, ``Incentive-driven task
  allocation for collaborative edge computing in industrial internet of
  things,'' \emph{IEEE Internet of Things Journal}, vol.~9, no.~1, pp.
  706--718, 2021.

\bibitem{huynh2023artificial}
T.~Huynh-The, Q.-V. Pham, X.-Q. Pham, T.~T. Nguyen, Z.~Han, and D.-S. Kim,
  ``Artificial intelligence for the metaverse: A survey,'' \emph{Engineering
  Applications of Artificial Intelligence}, vol. 117, p. 105581, 2023.

\bibitem{thomason2022metaverse}
J.~Thomason, ``Metaverse, token economies, and non-communicable diseases,''
  \emph{Global Health Journal}, vol.~6, no.~3, pp. 164--167, 2022.

\bibitem{gamefi}
\BIBentryALTinterwordspacing
``{GameFi},'' 2022. [Online]. Available: \url{https://gamefi.org/}
\BIBentrySTDinterwordspacing

\bibitem{Sandbox}
\BIBentryALTinterwordspacing
``{The Sandbox Game},'' 2022. [Online]. Available:
  \url{https://www.sandbox.game/en/}
\BIBentrySTDinterwordspacing

\bibitem{zhao2020privacy}
Y.~Zhao, J.~Zhao, L.~Jiang, R.~Tan, D.~Niyato, Z.~Li, L.~Lyu, and Y.~Liu,
  ``Privacy-preserving blockchain-based federated learning for iot devices,''
  \emph{IEEE Internet of Things Journal}, vol.~8, no.~3, pp. 1817--1829, 2020.

\bibitem{yuan2021coopedge}
L.~Yuan, Q.~He, S.~Tan, B.~Li, J.~Yu, F.~Chen, H.~Jin, and Y.~Yang, ``Coopedge:
  A decentralized blockchain-based platform for cooperative edge computing,''
  in \emph{Proceedings of the Web Conference 2021}, 2021, pp. 2245--2257.

\bibitem{nguyen2022metachain}
C.~T. Nguyen, D.~T. Hoang, D.~N. Nguyen, and E.~Dutkiewicz, ``Metachain: A
  novel blockchain-based framework for metaverse applications,'' in \emph{2022
  IEEE 95th Vehicular Technology Conference:(VTC2022-Spring)}.\hskip 1em plus
  0.5em minus 0.4em\relax IEEE, 2022, pp. 1--5.

\bibitem{EthereumWallet}
\BIBentryALTinterwordspacing
``{ETHEREUM WALLETS},'' 2022. [Online]. Available:
  \url{https://ethereum.org/en/wallets/}
\BIBentrySTDinterwordspacing

\bibitem{metamask}
\BIBentryALTinterwordspacing
``{MetaMask},'' 2022. [Online]. Available: \url{https://metamask.io/}
\BIBentrySTDinterwordspacing

\bibitem{WalletReview}
S.~Suratkar, M.~Shirole, and S.~Bhirud, ``Cryptocurrency wallet: A review,'' in
  \emph{2020 4th International Conference on Computer, Communication and Signal
  Processing (ICCCSP)}, 2020, pp. 1--7.

\bibitem{argent}
\BIBentryALTinterwordspacing
``{Argent Wallet},'' 2022. [Online]. Available: \url{https://www.argent.xyz/}
\BIBentrySTDinterwordspacing

\bibitem{zksync}
\BIBentryALTinterwordspacing
``{zksync},'' 2022. [Online]. Available: \url{https://zksync.io/}
\BIBentrySTDinterwordspacing

\bibitem{defillama}
\BIBentryALTinterwordspacing
``{DefiLlama},'' 2022. [Online]. Available:
  \url{https://defillama.com/categories}
\BIBentrySTDinterwordspacing

\bibitem{dexes}
\BIBentryALTinterwordspacing
``{Dexes},'' 2022. [Online]. Available:
  \url{https://defillama.com/protocols/Dexes}
\BIBentrySTDinterwordspacing

\bibitem{Lending}
\BIBentryALTinterwordspacing
``{Lending},'' 2022. [Online]. Available:
  \url{https://defillama.com/protocols/Lending}
\BIBentrySTDinterwordspacing

\bibitem{cdp}
\BIBentryALTinterwordspacing
``{CDP},'' 2022. [Online]. Available: \url{https://defillama.com/protocols/CDP}
\BIBentrySTDinterwordspacing

\bibitem{Liquid}
\BIBentryALTinterwordspacing
``{Liquid Staking},'' 2022. [Online]. Available:
  \url{https://defillama.com/protocols/Liquid%20Staking}
\BIBentrySTDinterwordspacing

\bibitem{Bridge}
\BIBentryALTinterwordspacing
``{Bridge},'' 2022. [Online]. Available:
  \url{https://defillama.com/protocols/Bridge}
\BIBentrySTDinterwordspacing

\bibitem{Yield}
\BIBentryALTinterwordspacing
``{Yield},'' 2022. [Online]. Available:
  \url{https://defillama.com/protocols/Yield}
\BIBentrySTDinterwordspacing

\bibitem{Services}
\BIBentryALTinterwordspacing
``{Services},'' 2022. [Online]. Available:
  \url{https://defillama.com/protocols/Services}
\BIBentrySTDinterwordspacing

\bibitem{Derivatives}
\BIBentryALTinterwordspacing
``{Derivatives},'' 2022. [Online]. Available:
  \url{https://defillama.com/protocols/Derivatives}
\BIBentrySTDinterwordspacing

\bibitem{Yield_Aggregator}
\BIBentryALTinterwordspacing
``{Yield Aggregator},'' 2022. [Online]. Available:
  \url{https://defillama.com/protocols/Yield%20Aggregator}
\BIBentrySTDinterwordspacing

\bibitem{Algo-Stables}
\BIBentryALTinterwordspacing
``{Algo-Stables},'' 2022. [Online]. Available:
  \url{https://defillama.com/protocols/Algo-Stables}
\BIBentrySTDinterwordspacing

\bibitem{Cross_Chain}
\BIBentryALTinterwordspacing
``{Cross Chain},'' 2022. [Online]. Available:
  \url{https://defillama.com/protocols/Cross%20Chain}
\BIBentrySTDinterwordspacing

\bibitem{Synthetics}
\BIBentryALTinterwordspacing
``{Synthetics},'' 2022. [Online]. Available:
  \url{https://defillama.com/protocols/Synthetics}
\BIBentrySTDinterwordspacing

\bibitem{Launchpad}
\BIBentryALTinterwordspacing
``{Launchpad},'' 2022. [Online]. Available:
  \url{https://defillama.com/protocols/Launchpad}
\BIBentrySTDinterwordspacing

\bibitem{Reserve_Currency}
\BIBentryALTinterwordspacing
``{Reserve Currency},'' 2022. [Online]. Available:
  \url{https://defillama.com/protocols/Reserve%20Currency}
\BIBentrySTDinterwordspacing

\bibitem{Insurance}
\BIBentryALTinterwordspacing
``{Insurance},'' 2022. [Online]. Available:
  \url{https://defillama.com/protocols/Insurance}
\BIBentrySTDinterwordspacing

\bibitem{Uniswap}
\BIBentryALTinterwordspacing
``{Uniswap},'' 2022. [Online]. Available: \url{https://docs.uniswap.org/}
\BIBentrySTDinterwordspacing

\bibitem{curve}
\BIBentryALTinterwordspacing
``{Curve},'' 2022. [Online]. Available:
  \url{https://medium.com/imtoken/defi-what-is-curve-and-how-do-i-use-it-4da29b2743ca}
\BIBentrySTDinterwordspacing

\bibitem{Saddle}
\BIBentryALTinterwordspacing
``{Saddle},'' 2022. [Online]. Available: \url{https://docs.saddle.finance/}
\BIBentrySTDinterwordspacing

\bibitem{Compound}
\BIBentryALTinterwordspacing
``{Compound},'' 2022. [Online]. Available: \url{https://docs.compound.finance/}
\BIBentrySTDinterwordspacing

\bibitem{Aave}
\BIBentryALTinterwordspacing
``{Aave},'' 2022. [Online]. Available: \url{https://docs.aave.com/}
\BIBentrySTDinterwordspacing

\bibitem{markerDAO}
\BIBentryALTinterwordspacing
``{MarkerDAO},'' 2022. [Online]. Available: \url{makerdao.com}
\BIBentrySTDinterwordspacing

\bibitem{2ndlife}
\BIBentryALTinterwordspacing
``{The Second Life},'' 2022. [Online]. Available: \url{https://secondlife.com/}
\BIBentrySTDinterwordspacing

\bibitem{Tilia}
\BIBentryALTinterwordspacing
``{Tilia},'' 2022. [Online]. Available: \url{https://www.tilia.io/}
\BIBentrySTDinterwordspacing

\bibitem{Decentraland}
\BIBentryALTinterwordspacing
``{Decentraland},'' 2022. [Online]. Available: \url{https://decentraland.org/}
\BIBentrySTDinterwordspacing

\bibitem{Highstreet}
\BIBentryALTinterwordspacing
``{Highstreet Market},'' 2022. [Online]. Available:
  \url{https://www.highstreet.market/}
\BIBentrySTDinterwordspacing

\bibitem{Polygon}
\BIBentryALTinterwordspacing
``{Polygon},'' 2022. [Online]. Available: \url{https://polygon.technology/}
\BIBentrySTDinterwordspacing

\bibitem{plasma}
\BIBentryALTinterwordspacing
``{Plasma},'' 2022. [Online]. Available:
  \url{https://ethereum.org/en/developers/docs/scaling/plasma/}
\BIBentrySTDinterwordspacing

\bibitem{Polygon_Pos}
\BIBentryALTinterwordspacing
``{Polygon Pos},'' 2022. [Online]. Available:
  \url{https://polygon.technology/solutions/polygon-pos}
\BIBentrySTDinterwordspacing

\bibitem{optimisticRollups}
\BIBentryALTinterwordspacing
``{Optimistic Rollups},'' 2022. [Online]. Available:
  \url{https://ethereum.org/en/developers/docs/scaling/optimistic-rollups/}
\BIBentrySTDinterwordspacing

\bibitem{coinmarketcap}
\BIBentryALTinterwordspacing
``{Coinmarketcap},'' 2022. [Online]. Available:
  \url{https://coinmarketcap.com/}
\BIBentrySTDinterwordspacing

\bibitem{multichain}
\BIBentryALTinterwordspacing
``{Multi-chain future likely as Ethereum’s DeFi dominance declines Bloomberg
  Professional Services},'' 2022. [Online]. Available:
  \url{https://www.bloomberg.com/professional/blog/multi-chain-future-likely-as-ethereums-defi-dominancede-clines/}
\BIBentrySTDinterwordspacing

\bibitem{multichainapproach}
\BIBentryALTinterwordspacing
``{A multichain approach is the future of the blockchain industry},'' 2022.
  [Online]. Available:
  \url{https://cointelegraph.com/news/a-multichain-approach-is-the-future-of-theblockchain-industry}
\BIBentrySTDinterwordspacing

\bibitem{Polkadot}
\BIBentryALTinterwordspacing
A.~Takyar, ``{CROSS-CHAIN WEB3 APPLICATIONS ON POLKADOT},'' 2022. [Online].
  Available:
  \url{https://www.leewayhertz.com/cross-chain-web3-applications-on-polkadot/}
\BIBentrySTDinterwordspacing

\bibitem{zamyatinAZKMKK21}
A.~Zamyatin, M.~Al{-}Bassam, D.~Zindros, E.~Kokoris{-}Kogias,
  P.~Moreno{-}Sanchez, A.~Kiayias, and W.~J. Knottenbelt, ``Sok: Communication
  across distributed ledgers,'' in \emph{Financial Cryptography and Data
  Security}, vol. 12675, 2021, pp. 3--36.

\bibitem{coinbase}
\BIBentryALTinterwordspacing
``Coinbase,'' 2022. [Online]. Available: \url{https://www.coinbase.com/}
\BIBentrySTDinterwordspacing

\bibitem{Tokrex}
\BIBentryALTinterwordspacing
{Mayer Christoph,Mai Jesse N, and Tom M}, ``{Tokrex},'' 2017. [Online].
  Available: \url{www.tokrex.org}
\BIBentrySTDinterwordspacing

\bibitem{corda}
\BIBentryALTinterwordspacing
``{Corda},'' 2022. [Online]. Available: \url{https://corda.net/}
\BIBentrySTDinterwordspacing

\bibitem{BTCB}
\BIBentryALTinterwordspacing
``{BTCB},'' 2022. [Online]. Available:
  \url{https://www.binance.com/en/how-to-buy/bitcoin-bep2}
\BIBentrySTDinterwordspacing

\bibitem{HBTC}
\BIBentryALTinterwordspacing
``{HBTC},'' 2022. [Online]. Available:
  \url{https://www.huobi.com/support/en-us/detail/900000196603/}
\BIBentrySTDinterwordspacing

\bibitem{tBTC}
\BIBentryALTinterwordspacing
``{tBTC},'' 2022. [Online]. Available:
  \url{https://docs.keep.network/tbtc/index.pdf}
\BIBentrySTDinterwordspacing

\bibitem{ren}
\BIBentryALTinterwordspacing
``{ren},'' 2022. [Online]. Available: \url{https://renproject.io/}
\BIBentrySTDinterwordspacing

\bibitem{DeCus}
\BIBentryALTinterwordspacing
``{DeCus},'' 2022. [Online]. Available: \url{https://docs.decus.io/mechanism}
\BIBentrySTDinterwordspacing

\bibitem{HopExchange}
\BIBentryALTinterwordspacing
``{Hop Exchange},'' 2022. [Online]. Available:
  \url{https://hop.exchange/whitepaper.pdf}
\BIBentrySTDinterwordspacing

\bibitem{Hyphen}
\BIBentryALTinterwordspacing
``{Hyphen},'' 2022. [Online]. Available:
  \url{https://docs.biconomy.io/products/hyphen-instant-cross-chain-transfers}
\BIBentrySTDinterwordspacing

\bibitem{DegateBridge}
\BIBentryALTinterwordspacing
``{Degate Bridge},'' 2022. [Online]. Available:
  \url{https://common-resource.degate.com/doc/whitepaper_cn.pdf}
\BIBentrySTDinterwordspacing

\bibitem{herlihy2018atomic}
M.~Herlihy, ``Atomic cross-chain swaps,'' in \emph{Proceedings of the 2018 ACM
  symposium on principles of distributed computing}, 2018, pp. 245--254.

\bibitem{WBTC}
\BIBentryALTinterwordspacing
``{WBTC},'' 2022. [Online]. Available:
  \url{https://wbtc.network/assets/wrapped-tokens-whitepaper.pdf}
\BIBentrySTDinterwordspacing

\bibitem{cBridge}
\BIBentryALTinterwordspacing
``{cBridge},'' 2022. [Online]. Available:
  \url{https://cbridge-docs.celer.network/#/}
\BIBentrySTDinterwordspacing

\bibitem{LightningNetwork}
\BIBentryALTinterwordspacing
``{Lightning Network},'' 2022. [Online]. Available:
  \url{https://lightning.network/}
\BIBentrySTDinterwordspacing

\bibitem{li2021sepow}
T.~Li, M.~Wang, Z.~Deng, and D.~Liu, ``Sepow: Secure and efficient proof of
  work sidechains,'' in \emph{International Conference on Algorithms and
  Architectures for Parallel Processing}.\hskip 1em plus 0.5em minus
  0.4em\relax Springer, 2021, pp. 376--396.

\bibitem{ZcashXCAT}
\BIBentryALTinterwordspacing
``{Zcash XCAT},'' 2022. [Online]. Available:
  \url{https://github.com/zcash/zcash/projects/4}
\BIBentrySTDinterwordspacing

\bibitem{Interledger}
\BIBentryALTinterwordspacing
``{Interledger},'' 2022. [Online]. Available:
  \url{https://interledger.org/interledger.pdf}
\BIBentrySTDinterwordspacing

\bibitem{btcrelay}
\BIBentryALTinterwordspacing
``{Btcrelay},'' 2019. [Online]. Available:
  \url{https://github.com/ethereum/btcrelay}
\BIBentrySTDinterwordspacing

\bibitem{zamyatin2019xclaim}
A.~Zamyatin, D.~Harz, J.~Lind, P.~Panayiotou, A.~Gervais, and W.~Knottenbelt,
  ``Xclaim: Trustless, interoperable, cryptocurrency-backed assets,'' in
  \emph{IEEE Symposium on Security and Privacy}.\hskip 1em plus 0.5em minus
  0.4em\relax IEEE, 2019, pp. 193--210.

\bibitem{xie2022zkbridge}
T.~Xie, J.~Zhang, Z.~Cheng, F.~Zhang, Y.~Zhang, Y.~Jia, D.~Boneh, and D.~Song,
  ``zkbridge: Trustless cross-chain bridges made practical,'' in \emph{ACM
  Conference on Computer and Communications Security}, 2022, pp. 3003--3017.

\bibitem{cosmos}
\BIBentryALTinterwordspacing
``{Cosmos},'' 2022. [Online]. Available: \url{https://cosmos.network/}
\BIBentrySTDinterwordspacing

\bibitem{ghosh2021leveraging}
B.~C. Ghosh, T.~Bhartia, S.~K. Addya, and S.~Chakraborty, ``Leveraging
  public-private blockchain interoperability for closed consortium
  interfacing,'' in \emph{IEEE Conference on Computer Communications}.\hskip
  1em plus 0.5em minus 0.4em\relax IEEE, 2021, pp. 1--10.

\bibitem{Ballardini2019AIgeneratedCA}
R.~M. Ballardini, ``Ai-generated content: authorship and inventorship in the
  age of artificial intelligence,'' \emph{Online Distribution of Content in the
  EU}, 2019.

\bibitem{heller2016avatars}
L.~Heller and L.~Goodman, ``What do avatars want now? posthuman embodiment and
  the technological sublime,'' in \emph{2016 22nd International Conference on
  Virtual System \& Multimedia (VSMM)}.\hskip 1em plus 0.5em minus 0.4em\relax
  IEEE, 2016, pp. 1--4.

\bibitem{genay2021being}
A.~Genay, A.~L{\'e}cuyer, and M.~Hachet, ``Being an avatar “for real”: a
  survey on virtual embodiment in augmented reality,'' \emph{IEEE Transactions
  on Visualization and Computer Graphics}, vol.~28, no.~12, pp. 5071--5090,
  2021.

\bibitem{liao2021digital}
S.~Liao, J.~Wu, A.~K. Bashir, W.~Yang, J.~Li, and U.~Tariq, ``Digital twin
  consensus for blockchain-enabled intelligent transportation systems in smart
  cities,'' \emph{IEEE Transactions on Intelligent Transportation Systems},
  2021.

\bibitem{yang2022secure}
K.~Yang, Z.~Zhang, Y.~Tian, and J.~Ma, ``A secure authentication framework to
  guarantee the traceability of avatars in metaverse,'' \emph{arXiv preprint
  arXiv:2209.08893}, 2022.

\bibitem{tian2021enabling}
H.~Tian, K.~Xue, X.~Luo, S.~Li, J.~Xu, J.~Liu, J.~Zhao, and D.~S. Wei,
  ``Enabling cross-chain transactions: A decentralized cryptocurrency exchange
  protocol,'' \emph{IEEE Transactions on Information Forensics and Security
  (TIFS)}, vol.~16, pp. 3928--3941, 2021.

\bibitem{Wu2020who}
J.~Wu, Q.~Yuan, D.~Lin, W.~You, W.~Chen, C.~Chen, and Z.~Zheng, ``{Who are the
  phishers? Phishing scam detection on ethereum via network embedding},''
  \emph{IEEE Trans. Syst. Man, Cybern. Syst. (TSMC)}, vol.~52, no.~2, pp.
  1156--1166, Feb. 2022.

\bibitem{Chen2018Detecting}
W.~Chen, Z.~Zheng, J.~Cui, E.~Ngai, P.~Zheng, and Y.~Zhou, ``{Detecting Ponzi
  schemes on Ethereum: Towards healthier blockchain technology},'' in
  \emph{Proc. of WWW}.\hskip 1em plus 0.5em minus 0.4em\relax ACM, Apr. 2018,
  pp. 1409--1418.

\bibitem{CrossChainMetaverse}
\BIBentryALTinterwordspacing
``{Web3 Engine Leads the Cross-Chain Future of the Metaverse},'' 2022.
  [Online]. Available:
  \url{https://medium.com/rangersprotocol/web3-engine-leads-the-cross-chain-future-of-the-metaverse-3b2ba8b02d5e#:~:text=Cross-chain\%20is\%20the\%20key\%20to\%20the\%20multi-chain\%20future,to\%20realize\%20the\%20cross-chain\%20future\%20of\%20the\%20Metaverse.}
\BIBentrySTDinterwordspacing

\bibitem{Liu2021DAO}
L.~Liu, S.~Zhou, H.~Huang, and Z.~Zheng, ``From technology to society: An
  overview of blockchain-based {DAO},'' \emph{{IEEE Open Journal of the
  Computer Society (OJ-CS)}}, vol.~2, pp. 204--215, 2021.

\end{thebibliography}

\begin{IEEEbiography}[{\includegraphics[width=1in,height=1.25in,clip,keepaspectratio]{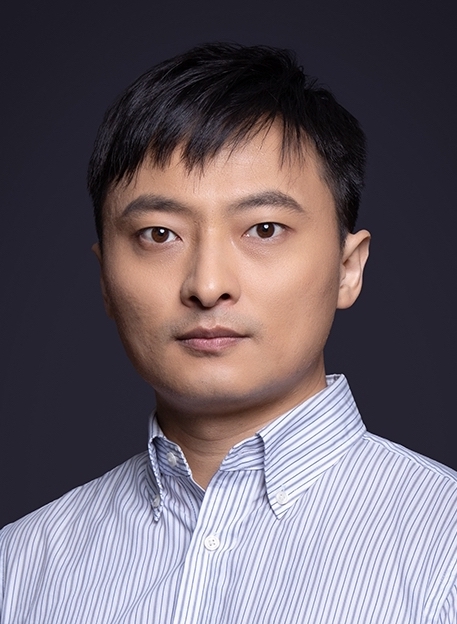}}]{Huawei~Huang}
 (M'16, SM'22) received his Ph.D in Computer Science and Engineering from the University of Aizu, Japan. His research interests mainly include blockchains. He received the best paper award in TrustCom-2016. He has served as a visiting scholar at the Hong Kong Polytechnic University and as an assistant professor at Kyoto University, Japan. He is currently with 
 the School of Engineering, Sun Yat-Sen University, China.
\end{IEEEbiography}

\end{document}